\PassOptionsToPackage{svgnames}{xcolor}

\documentclass[sigconf,screen,nonacm]{acmart}

\AtBeginDocument{%
  }

\usepackage{graphicx}
\usepackage{hyperref}
\usepackage{hyperxmp}
\usepackage{multirow}

\usepackage{booktabs}   
\usepackage{subcaption} 
\usepackage{xspace}
\usepackage{pifont}
\usepackage{stfloats}

\usepackage{fancybox}
\usepackage{stmaryrd}

\usepackage{listings-rust} 
\usepackage{listings-scribble} 
\newcommand{\xmark}{\ding{55}} 
\newcommand{\cmark}{\ding{51}} 
\usepackage{cleveref} 
\usepackage{amsthm}
\usepackage{mathtools}
\usepackage{amsmath}
\usepackage{textcomp}
\usepackage{tabularx}
\usepackage{xr}
\usepackage{makecell}
\usepackage{xspace}
\usepackage{tikz}
\usepackage[clock]{ifsym}
\usepackage{pifont}
\usetikzlibrary{
  arrows,
  automata,
  calc,
  fit,
  positioning,
  shapes,
  arrows.meta,
  shapes.multipart,
  tikzmark,
  backgrounds
}
\usepackage{algorithm}
\usepackage[noend]{algpseudocode}
\usepackage[most]{tcolorbox}
\tcbuselibrary{listings, breakable}
\tcbset{
    colback=gray!2,
    colframe=black,
    boxrule=0.4pt,
    arc=1mm,
    left=2pt,
    right=2pt,
    top=2pt,
    bottom=2pt,
    enhanced
  }

\usepackage[most]{tcolorbox}
\usepackage[table]{xcolor}
\usepackage[nomargin,inline,index,%
  status=draft 
]{fixme} 
\fxusetheme{colorsig}
\FXRegisterAuthor{fxAS}{anfxAS}{AS}
\FXRegisterAuthor{fxAB}{anfxAB}{AB}
\FXRegisterAuthor{fxNY}{anfxNY}{NY}
\FXRegisterAuthor{fxFZ}{anfxFZ}{FZ}
\FXRegisterAuthor{fxPH}{anfxPH}{PH}
\FXRegisterAuthor{fxYL}{anfxYL}{YL}

\makeatletter
\DeclareFontFamily{OMX}{MnSymbolE}{}
\DeclareSymbolFont{MnLargeSymbols}{OMX}{MnSymbolE}{m}{n}
\SetSymbolFont{MnLargeSymbols}{bold}{OMX}{MnSymbolE}{b}{n}
\DeclareFontShape{OMX}{MnSymbolE}{m}{n}{
    <-6>  MnSymbolE5
   <6-7>  MnSymbolE6
   <7-8>  MnSymbolE7
   <8-9>  MnSymbolE8
   <9-10> MnSymbolE9
  <10-12> MnSymbolE10
  <12->   MnSymbolE12
}{}
\DeclareFontShape{OMX}{MnSymbolE}{b}{n}{
    <-6>  MnSymbolE-Bold5
   <6-7>  MnSymbolE-Bold6
   <7-8>  MnSymbolE-Bold7
   <8-9>  MnSymbolE-Bold8
   <9-10> MnSymbolE-Bold9
  <10-12> MnSymbolE-Bold10
  <12->   MnSymbolE-Bold12
}{}

\let\llangle\@undefined
\let\rrangle\@undefined
\DeclareMathDelimiter{\llangle}{\mathopen}%
                     {MnLargeSymbols}{'164}{MnLargeSymbols}{'164}
\DeclareMathDelimiter{\rrangle}{\mathclose}%
                     {MnLargeSymbols}{'171}{MnLargeSymbols}{'171}
\makeatother

\usepackage[inline]{enumitem}%

\usepackage{nicefrac}

\usepackage{proof}

\usepackage[most]{tcolorbox}
\newtcolorbox{myframe}[1][]{
  enhanced,
  arc=0pt,
  outer arc=0pt,
  colback=white,
  boxrule=0.5pt,
  boxsep=0mm,
  left=1mm,
  right=1mm,
  top=0.5mm,
  bottom=0.5mm,
  #1
}

\lstdefinelanguage{Scribble}{%
  basicstyle=\footnotesize\ttfamily,
  stringstyle=\color{Blue},
  showstringspaces=false,
  keywords={nested,new,calls,and,as,at,by,catches,choice,continue,do,from,global,import,instantiates,interruptible,local,module,or,par,protocol,rec,role,sig,throws,to,type,with,int,aux,reliable,crash},
  morestring=[b]",
  morestring=[b]',
  morecomment=[l][\color{greencomments}]{//},
}

\lstdefinelanguage{nuScr}{%
  basicstyle=\footnotesize\ttfamily,
  stringstyle=\color{Blue},
  showstringspaces=false,
  keywords={
    nested,new,calls,and,as,at,by,catches,choice,continue,do,from,global,import,instantiates,interruptible,local,module,or,par,protocol,rec,role,sig,throws,to,type,with,int,aux,
    safe
  },
  morestring=[b]",
  morestring=[b]',
  morecomment=[l][\color{greencomments}]{//},
  morecomment=[s][\color{magenta}]{(*}{*)},
}

\lstdefinelanguage{effpi}{
  keywords=[1]{
    case,class,sealed,abstract,object,extends,type,def,val,if,else,new,var,match
  },
  keywords=[2]{
    InChan,OutChan,RecVar,Rec,Out,In,InErr,Loop,
  },
  keywords=[3]{
    rec,send,receive,receiveErr,eval,par,Channel
  },
  keywordstyle=[1]{\color{blue}},
  keywordstyle=[2]{\color{ImperialIris}}, 
  keywordstyle=[3]{\color{OliveGreen}},
  otherkeywords={=>,.type,<:,>>:},
  morecomment=[l][\color{darkgray}]{//},
}

\usepackage{stmaryrd}
\usepackage{arydshln}

\usepackage{xifthen}
\usepackage{xspace}

\usepackage{mathtools}

\usepackage{balance}

\usepackage{hyperref}
\hypersetup{hidelinks}

\definecolor{ImperialBlue}{HTML}{003E74}
\definecolor{ImperialDarkGreen}{HTML}{02893B}
\definecolor{ImperialTangerine}{HTML}{EC7300}
\definecolor{ImperialIris}{HTML}{751E66}

\definecolor{RYB1}{RGB}{141, 211, 199}
\definecolor{RYB2}{RGB}{255, 255, 179}
\definecolor{RYB3}{RGB}{190, 186, 218}
\definecolor{RYB4}{RGB}{251, 128, 114}
\definecolor{RYB5}{RGB}{128, 177, 211}
\definecolor{RYB6}{RGB}{253, 180, 98}
\definecolor{RYB7}{RGB}{179, 222, 105}

\usetikzlibrary{automata, positioning, arrows, calc, fit}
\tikzset{
  >=stealth,
  node distance=2cm,
  every state/.style={thick, fill=gray!10},
  initial text=$ $,
}
\usepackage{pgfplots}
\pgfplotscreateplotcyclelist{stackexchange}{
  {RYB1!50!black,fill=RYB1},
  {RYB4!50!black,fill=RYB4},
  {RYB3!50!black,fill=RYB3},
  {RYB2!50!black,fill=RYB2},
  {RYB5!50!black,fill=RYB5},
  {RYB6!50!black,fill=RYB6},
  {RYB7!50!black,fill=RYB7},
}
\pgfplotscreateplotcyclelist{stackexchangePlusOne}{
  {RYB4!50!black,fill=RYB4},
  {RYB3!50!black,fill=RYB3},
  {RYB2!50!black,fill=RYB2},
  {RYB5!50!black,fill=RYB5},
  {RYB6!50!black,fill=RYB6},
  {RYB7!50!black,fill=RYB7},
}
\pgfplotsset{
  compat=1.8,
  /pgfplots/bar cycle list/.style={/pgfplots/cycle list={%
    {brown!60!black,fill=brown!30!white,mark=none},
    {red,fill=red!30!white,mark=none},
    {blue,fill=blue!30!white,mark=none},
    {black,fill=gray,mark=none},
    }
  },
}
\newcolumntype{L}{>{$}l<{$}}
\newcolumntype{C}{>{$}c<{$}}
\newcolumntype{P}[1]{>{\centering\arraybackslash$}p{#1}<{$}}
\usepackage{multirow}

\usepackage[export]{adjustbox}

\Crefname{section}{\S\!}{\S\!}%
\Crefname{subsection}{\S\!}{\S\!}%
\Crefname{subsubsection}{\S\!}{\S\!}%
\Crefname{appendix}{Appendix \S\!}{Appendix \S\!}
\Crefname{definition}{Def.\@}{Defs.\@}%
\Crefname{figure}{Fig.\@}{Figs.\@}%
\Crefname{example}{Ex.\@}{Exs.\@}%
\Crefname{corollary}{Cor.\@}{Cors.\@}%
\Crefname{theorem}{Thm.\@}{Thms.\@}%
\Crefname{proposition}{Prop.\@}{Props.\@}%
\Crefname{lemma}{Lem.\@}{Lems.\@}
\Crefname{equation}{Eq.\@}{Eqs.\@}

\crefname{section}{\S\!}{\S\!}%
\crefname{subsection}{\S\!}{\S\!}%
\crefname{subsubsection}{\S\!}{\S\!}%
\crefname{appendix}{Appendix \S\!}{Appendix \S\!}
\crefname{definition}{Def.\@}{Defs.\@}%
\crefname{figure}{Fig.\@}{Figs.\@}%
\crefname{example}{Ex.\@}{Exs.\@}%
\crefname{corollary}{Cor.\@}{Cors.\@}%
\crefname{theorem}{Thm.\@}{Thms.\@}%
\crefname{proposition}{Prop.\@}{Props.\@}%
\crefname{lemma}{Lem.\@}{Lems.\@}
\crefname{equation}{Eq.\@}{Eqs.\@}

\usepackage{magicvariables}

\newif\ifdraft%
  \draftfalse%
  \drafttrue%

\newcommand{\ifempty}[3]{%
  \ifthenelse{\isempty{#1}}{#2}{#3}%
}%

\newtoggle{techreport}%

\newtoggle{cruft}%
\newcommand{\notImpliedBy}{\mathrel{{\kern .5em}{\not{\kern -1.2em}\impliedby}}}%
\newcommand{\coloncolonequals}{\Coloneqq}
\newcommand{\bnfdef}{\coloncolonequals}%
\newcommand{\bnfsep}{\mathbin{\;\big|\;}}%

\newcommand{\theTool}[0]{\textsc{Syntropy}\xspace}
\newcommand{\trainPhase}[0]{\textsc{Syntropy-Train}\xspace}
\newcommand{\inferencePhase}[0]{\textsc{Syntropy-Gen}\xspace}

\newcommand{\Scala}[0]{\textsc{Scala}\xspace}
\newcommand{\Rust}[0]{\textsc{Rust}\xspace}
\newcommand{\Go}[0]{\textsc{Go}\xspace}
\newcommand{\Java}[0]{\textsc{Java}\xspace}

\newcommand{\Python}[0]{\textsc{Python}\xspace}

\newcommand{\RefA}[0]{\textsc{RefA}\xspace}
\newcommand{\RefB}[0]{\textsc{RefB}\xspace}
\newcommand{\RefIn}[0]{\textsc{RefIn}\xspace}
\newcommand{\RefOut}[0]{\textsc{RefOut}\xspace}
\newcommand{\Unfold}[0]{\textsc{Unfold}\xspace}
\newcommand{\Identity}[0]{\textsc{Identity}\xspace}

\def\eg{e.g.\@\xspace}%
\def\ie{i.e.\@\xspace}%
\definecolor{ruleColor}{rgb}{0.1, 0.3, 0.1}
\definecolor{hlColor}{rgb}{0.65, 1.0, 0.65}%

\newcommand{\runtime}[2][\runtimeColour]{\mathchoice%
  {\setlength{\fboxsep}{0pt}\colorbox{#1}{$\displaystyle#2$}}%
  {\setlength{\fboxsep}{0pt}\colorbox{#1}{$\textstyle#2$}}%
  {\setlength{\fboxsep}{0pt}\colorbox{#1}{$\scriptstyle#2$}}%
  {\setlength{\fboxsep}{0pt}\colorbox{#1}{$\scriptscriptstyle#2$}}}%

\newcommand{\lbbar}{\{\kern-0.2em|}
\newcommand{\rbbar}{|\kern-0.2em\}}

\newcommand{\MPST}[0]{{\textsf{MPST}}\xspace}%

\definecolor{tyColorCustom}{rgb}{0.0, 0.0, 0.85}%
\newcommand{\muCol}[1]{{\color{red}#1}}%
\newcommand{\muWordEmpty}[1][]{\muCol{\epsilon}}%
{\centerline{\bf --- Begin Copied From Previous Paper ---} \hrule}%
{\hrule \centerline{\bf --- End Copied From Previous Paper ---}}%

{\centerline{\bf --- Begin Discussion\ifempty{#1}{}{: {#1}} ---}
  \hrule\vspace{1mm}}%
{\hrule\vspace{1mm}\centerline{\bf --- End Discussion ---}}%

\newtcolorbox{cross}{blank,breakable,parbox=false,
  overlay={\draw[red,line width=5pt] (interior.south west)--(interior.north east);
    \draw[red,line width=5pt] (interior.north west)--(interior.south east);}}

\definecolor{roleColor}{rgb}{0.5, 0.0, 0.0}
\newcommand{\roleCol}[1]{{\color{Teal}#1}}%
\newcommand{\roleSet}{\roleCol{\mathcal{R}}}%
\newcommand{\roleFmt}[1]{\ensuremath{{\boldsymbol{\roleCol{\mathtt{#1}}}}}\xspace}%

\newcommand{\roleP}[1][]{%
  \ifempty{#1}{{\color{roleColor}\roleFmt{p}}}{{\color{roleColor}\roleFmt{p}_{#1}}}%
}%
\newcommand{\rolePi}[1][]{%
  \ifempty{#1}{{\color{roleColor}\roleFmt{p'}}}{{\color{roleColor}\roleFmt{p'}_{#1}}}%
}%

\newcommand{\roleQ}[1][]{%
  \ifempty{#1}{{\color{roleColor}\roleFmt{q}}}{{\color{roleColor}\roleFmt{q}_{#1}}}%
}%
\newcommand{\roleR}[1][]{%
  \ifempty{#1}{{\color{roleColor}\roleFmt{r}}}{{\color{roleColor}\roleFmt{r}_{\!#1}}}%
}%
\definecolor{gtColor}{rgb}{0.43, 0.21, 0.1}
\newcommand{\gtFmt}[1]{\ensuremath{{\color{gtColor}#1}}\xspace}%
\newcommand{\gtMsgFmt}[1]{\gtFmt{\labFmt{#1}}}%
\newcommand{\labFmt}[2][]{\ensuremath{\ifempty{#1}{\mathtt{#2}}{\mathtt{#2}\textsubscript{#1}}}\xspace}%

\definecolor{stColor}{rgb}{0, 0, 0.9}
\newcommand{\stFmt}[1]{\ensuremath{{\color{stColor}#1}}\xspace}%

\newcommand{\stChoice}[2]{\stLabFmt{#1}\ifempty{#2}{}{\stFmt{({#2})}}}%

\newcommand{\stSeq}{\mathbin{\!\stFmt{.}\!}}%
\newcommand{\stRec}[2]{\stFmt{\mu{#1}.{#2}}}%
\newcommand{\stEnd}{\stFmt{\mathbf{end}}}%

\newcommand{\stLabFmt}[1]{\stFmt{\labFmt{#1}}}%
\newcommand{\stLab}[1][]{%
  \ifempty{#1}{\stLabFmt{m}}{\stLabFmt{m}_{{\color{stColor}#1}}}
}%
\newcommand{\stLabi}[1][]{%
  \ifempty{#1}{\stLabFmt{m'}}{\stLabFmt{m'}_{{\color{stColor}#1}}}
}%

\newcommand{\trainLab}[1][]{%
  \ifempty{#1}{\mpLabFmt{m}}{\mpLabFmt{l}_{{\color{mpColor}#1}}}
}%

\newcommand{\stS}[1][]{\stFmt{\ifempty{#1}{S}{S_{#1}}}}%

\newcommand{\stT}[1][]{\stFmt{\ifempty{#1}{T}{T_{#1}}}}%
\newcommand{\stTi}[1][]{\stFmt{\ifempty{#1}{T'}{T'_{#1}}}}%
\newcommand{\stRecVarBase}{\stFmt{\mathbf{t}}}%
\newcommand{\stRecVar}[1][]{\stFmt{\ifempty{#1}{\stRecVarBase}{\stRecVarBase_{#1}}}}%
\newcommand{\stSub}{\mathrel{\stFmt{\leqslant}}}%
\newcommand{\stRefinement}{\mathrel{\stFmt{\lesssim}}}%

\definecolor{mpColor}{rgb}{0, 0, 0}
\newcommand{\mpFmt}[1]{{\color{mpColor}#1}}%

\newcommand{\mpLabFmt}[1]{\mpFmt{\labFmt{#1}}}%

\makeVars{mpE}{mpFmt}{e}

\newtoggle{full}
\settoggle{full}{false}
\iftoggle{full}
{
	
}{
	
}

\makeatletter
\newcommand{\iftoggleverb}[1]{%
  \ifcsdef{etb@tgl@#1}
    {\csname etb@tgl@#1\endcsname\iftrue\iffalse}
    {\etb@noglobal\etb@err@notoggle{#1}\iffalse}%
}
\makeatother

\begin{document}

\title{Specification-Guided Synthesis of Deadlock-Free Communication Protocol Refinements with Large Language Models}

\author{Yang Li}
\orcid{0009-0004-0478-9608}
\affiliation{%
  \institution{University of Oxford}
  \department{Computer Science}
  \city{Oxford}
  \country{United Kingdom}
}
\email{yang.li@wolfson.ox.ac.uk}

\author{Ping Hou}
\orcid{0000-0001-6899-9971}
\affiliation{%
  \institution{University of Oxford}
  \department{Computer Science}
  \city{Oxford}
  \country{United Kingdom}
}
\email{ping.hou@cs.ox.ac.uk}

\author{Nobuko Yoshida}
\orcid{0000-0002-3925-8557}
\affiliation{%
  \institution{University of Oxford}
  \department{Computer Science}
  \city{Oxford}
  \country{United Kingdom}
}
\email{nobuko.yoshida@cs.ox.ac.uk}


\begin{abstract}

Ensuring \emph{behavioural correctness} in \emph{communication protocols} is a central challenge in distributed software systems, as subtle inconsistencies can lead to deadlocks. In such settings, \emph{protocol refinement} -- the safe substitution of a protocol that preserves correctness and compatibility with other components -- 
is essential. 
Large language models (LLMs) have demonstrated strong capabilities in code generation and program synthesis, yet lack mechanisms to reliably produce outputs with correct behaviour. Formal specification approaches,  such as \emph{multiparty session types} (\MPST),  offer rigorous guarantees, including deadlock freedom, but provide limited support for automatically constructing protocol refinements. In this paper, we present \theTool, a framework for synthesising protocol refinements guided by \MPST specifications and LLMs. It  incorporates refinement constraints directly into the generation process, ensuring the generated variants satisfy these guarantees.  
Our comprehensive evaluation indicates that \theTool achieves 95.6\%--99.5\% validity while maintaining high syntactic correctness, and produces diverse, non-trivial refinements across multiple LLMs.
\end{abstract}







\maketitle

\section{Introduction}
\label{sec:intro}

Modern software engineering is increasingly shaped by large language models (LLMs)~\cite{NIPS2017_3f5ee243},  
which are widely used to automate programming tasks such as code generation~\cite{DBLP:journals/corr/abs-2107-03374}, transformation~\cite{DBLP:journals/corr/abs-2308-12950}, and program synthesis~\cite{DBLP:journals/corr/abs-2108-07732}. Recent work~\cite{DBLP:journals/corr/abs-2308-12950,DBLP:journals/pacmpl/MundlerHWSSV25,DBLP:journals/pacmpl/NagyZPD26} shows that LLMs can produce syntactically correct and functionally useful code across diverse tasks, including programs that satisfy certain semantic properties such as type safety, indicating their potential to support complex software development workflows.

Despite these advances, a key limitation remains: LLM-generated artefacts do not provide behavioural guarantees by construction, especially in systems with complex interactions. This is particularly critical  in distributed software systems, from microservices~\cite{DBLP:books/sp/17/DragoniGLMMMS17} to  cyber-physical systems~\cite{4519604}, where reliability depends on consistent communication between components as well as local functionality.  Subtle  interaction errors, 
such as incorrect message ordering or unintended cyclic dependencies, can lead to failures, including \emph{deadlocks}. Ensuring that LLM-generated or modified programs preserve reliable communication remains an open challenge. 

Such systems rely on \emph{communication protocols} to coordinate independently developed components. 
As systems evolve, protocols must be adapted to incorporate new functionality or modify interaction patterns. 
Even minor changes can have system-wide effects that are difficult to predict and diagnose. 
\emph{Protocol refinement} -- replacing a protocol with a behaviourally compatible alternative -- offers a principled approach to managing this evolution. Nevertheless, preserving the intended interaction properties during this process is non-trivial in practice. In many cases, refinements are constructed manually, making them error-prone and hard to validate. This motivates the following research question:
\vspace{-0.43em}
\begin{tcolorbox}[colback=gray!6, colframe=gray]
{\emph{How can protocol refinements be systematically synthesised while retaining  behavioural correctness?}}
\end{tcolorbox}
\vspace{-.4em}

 To address this question, we draw on specification techniques such as session types~\cite{HondaVK98CommProg}, in particular multiparty session types (\MPST)~\cite{HYC16,POPL19LessIsMore}, which are well established for describing communication protocols and reasoning about their behaviour.
 \MPST captures global interaction structures and enforcs properties such as communication safety and deadlock freedom.  
 Toolchains for \MPST have been developed for over 30 programming languages~\cite{DBLP:books/sp/24/Yoshida24}, facilitating  its adoption in software engineering.

 \MPST further defines refinement relations that allow protocols to be transformed into more flexible forms  while preserving these properties~\cite{GJPSY2019,GPPSY2023}. 
Although checking such relations can be automated~\cite{DBLP:conf/concur/BravettiCLYZ19,DBLP:conf/tacas/BocchiKM24,DBLP:conf/ppopp/CutnerYV22,DBLP:conf/concur/BocchiK0025}, generating variants that satisfy refinement constraints is not straightforward and, in some settings, \emph{undecidable}~\cite{DBLP:journals/iandc/BravettiCZ17,DBLP:conf/fossacs/LangeY17}, limiting the automation of protocol refinement.

To bridge this gap, we propose \theTool, the first framework combining LLMs with \MPST to support the synthesis of protocol refinements, to the best of our knowledge. Given an \MPST protocol, \theTool constructs protocol candidates based on structured representations of interaction behaviour and refinement objectives. The generation process is coupled with validation against refinement constraints, ensuring that only valid outputs are retained. This approach extends LLM-based generation beyond syntactic and local semantic aspects by incorporating protocol-level requirements, including deadlock freedom. By integrating generative models with specification-based verification, \theTool explores the space of valid protocol refinements and produces varied and reliable alternatives that are difficult to construct manually.

We evaluate \theTool on two datasets, comprising protocols derived from the literature and synthetic benchmarks, using multiple LLMs of varying sizes: three 7B code models, a general-purpose 7B model, and a 32B model. \theTool attains 95.6\%--99.5\% validity across all models, while maintaining strong syntactic correctness (95.4\%--98.1\%). Furthermore, it produces multiple distinct refinements for each specification, demonstrating its ability to identify alternative protocol formulations, rather than trivial variations. Ablation studies indicate that both specification guidance and 
constraint-based validation are essential for attaining high validity and diversity.  

We additionally compare \theTool with frontier language models, demonstrating that, although frontier language models can synthesise valid protocol refinements, their coverage across protocol specifications remains limited. This highlights the need for the proposed approach, which enables comprehensive protocol refinement while supporting reproducible evaluation and local deployment with fine-tuned open models.


The contributions of this paper are as follows:
\begin{itemize}[leftmargin=9pt,itemsep=1.5pt] 
\item {\bf LLM Generation with Behavioural Guarantees.} 
We propose a novel approach  to enable LLMs to synthesise \MPST protocol refinements with guaranteed behavioural correctness.

\item {\bf Specification-Guided Protocol Refinement.} 
We introduce a systematic encoding of \MPST specifications that guides and constrains LLM-based generation of protocol refinements.


\item {\bf Constraint-Integrated Generation.}  
We design a two-level generation workflow that incorporates constraint validation into the synthesis via prefix filtering and subsequent verification.


\item {\bf \theTool Framework and Empirical Evaluation.} 
We implement \theTool and evaluate it across multiple LLMs and datasets, demonstrating high validity and generating structurally distinct and non-superficial protocol refinements.

\end{itemize}

\begin{figure*}[!t]
    \centering
    \begin{minipage}[t]{0.44\textwidth}
        \vspace{0pt}
        \centering
        \includegraphics[width=\linewidth]{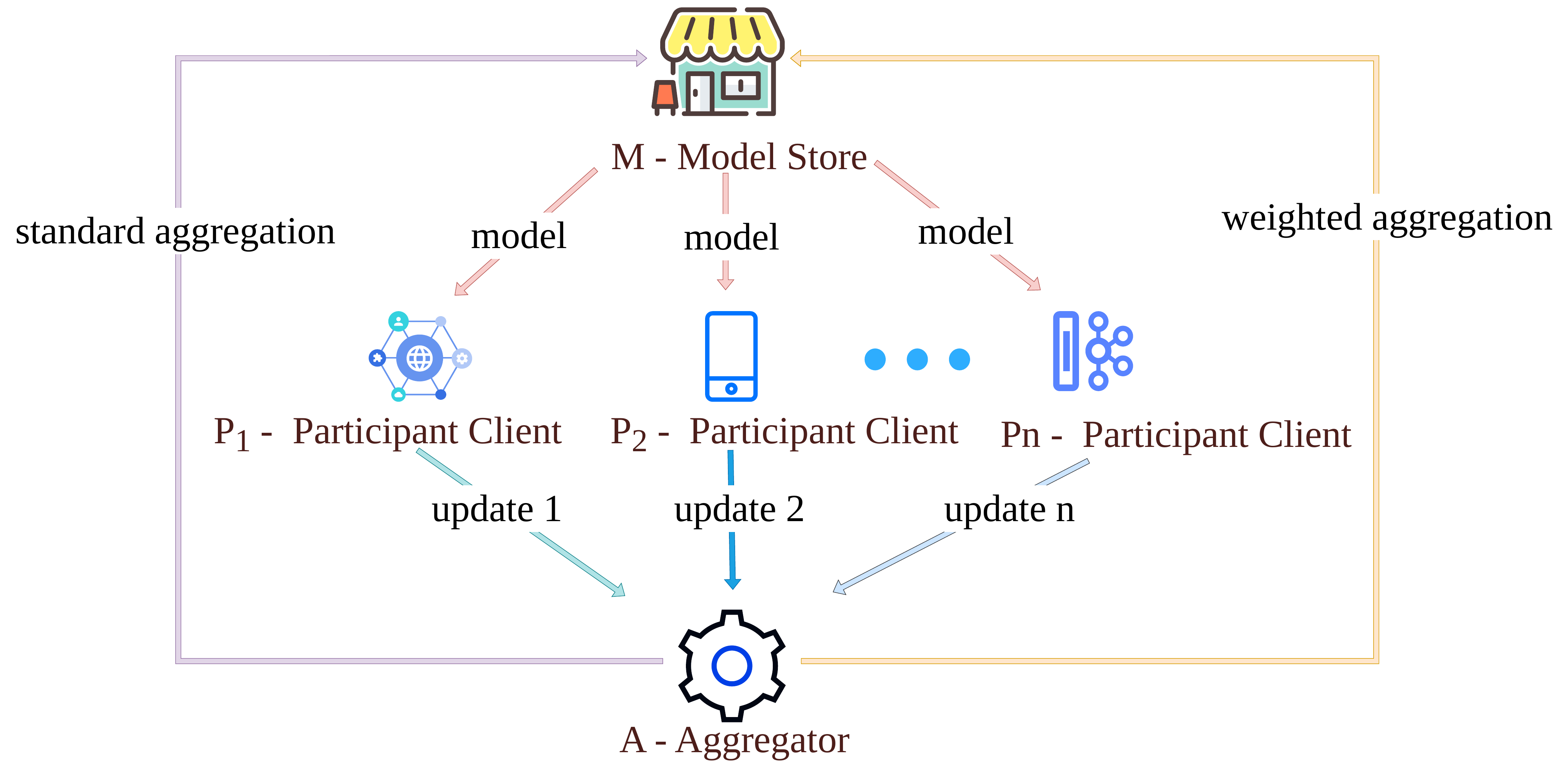}
        \caption{\small Federated learning protocol}
        \label{fig:federated_protocol}
    \end{minipage}
    \hfill
    \begin{minipage}[t]{0.55\textwidth}
        \vspace{0pt}
        \centering
        \begin{subfigure}[t]{0.45\linewidth}
          \centering 
            \begin{tikzpicture}[thick,baseline={([yshift=-.5ex]current bounding box.center)},level distance=2.5em, sibling distance=3em,style={font=\footnotesize}]
    \node {$\stFmt{\roleFmt{p}?}$}
    child { node {${\stLabFmt{upd}}_{\roleP}$} }
    child { node {$\stFmt{\roleFmt{q}?}$}
        child { node {${\stLabFmt{upd}}_{\roleQ}$} }
        child { node {$\stFmt{\roleFmt{m}!}$}
            child { node {$\stLabFmt{std}$} }
            child { node {$\stFmt{\roleFmt{p}?}$}
                child { node {\ldots} }
                child { node {\ldots} } }
            child { node {$\stLabFmt{wtd}$} }
            child { node {$\stFmt{\roleFmt{p}?}$}
                child { node {\ldots} }
                child { node {\ldots} } } } };
  \end{tikzpicture}
  \caption{Session tree $\stFmt{{\mathbb{T}}_{\roleFmt{a}}}$ for $\stT[\roleFmt{a}]$}
  \label{fig:session_tree_T1}
        \end{subfigure}
        \hspace{-2em}
        \begin{subfigure}[t]{0.45\linewidth}
            \centering
            \begin{tikzpicture}[thick,baseline={([yshift=-.5ex]current bounding box.center)},level distance=2.5em, sibling distance=3em,style={font=\footnotesize}]
    \node {$\stFmt{\roleFmt{q}?}$}
    child { node {${\stLabFmt{upd}}_{\roleQ}$} }
    child { node {$\stFmt{\roleFmt{p}?}$}
        child { node {${\stLabFmt{upd}}_{\roleP}$} }
        child { node {$\stFmt{\roleFmt{m}!}$}
            child { node {$\stLabFmt{std}$} }
            child { node {$\stFmt{\roleFmt{q}?}$}
                child { node {\ldots} }
                child { node {\ldots} } }
            child { node {$\stLabFmt{wtd}$} }
            child { node {$\stFmt{\roleFmt{q}?}$}
                child { node {\ldots} }
                child { node {\ldots} } } } };
  \end{tikzpicture}
  \caption{Session tree $\stFmt{{\mathbb{T}'}_{\!\!\roleFmt{a}}}$ for $\stTi[\roleFmt{a}]$ \kern-2em}
  \label{fig:session_tree_T_2}
        \end{subfigure}
        \caption{\small Session trees}
        \label{fig:session_tree}
    \end{minipage}
\end{figure*}

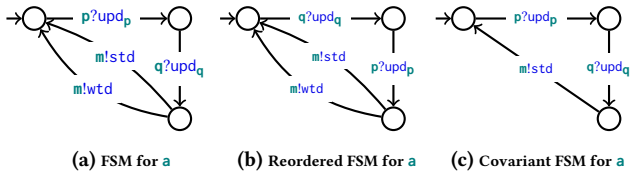
\begin{figure}[t]
  \centering

  \begin{subfigure}{.27\linewidth}
    \centering
    \begin{tikzpicture}[>=to, -to, thick, font=\scriptsize,
        initial left, initial distance=1.5ex, initial text={},
        state/.style={draw, circle, node distance=1cm and 1.6cm, inner sep=3pt},
        every path/.style={every node/.style={fill=white}}]

      \node[state,initial] (1) {};
      \node[state] (2) [right=of 1] {};
      \node[state] (3) [below=of 2] {};

      \path (1) edge node {$\stFmt{\roleFmt{p}?{\stLabFmt{upd}}_{\roleP}}$} (2);
      \path (2) edge node {$\stFmt{\roleFmt{q}?{\stLabFmt{upd}}_{\roleQ}}$} (3);
      \path (3) edge[bend right=15] node {$\stFmt{\roleFmt{m}!\stLabFmt{std}}$} (1);
      \path (3) edge[bend left=25] node {$\stFmt{\roleFmt{m}!\stLabFmt{wtd}}$} (1); 
    \end{tikzpicture}
    \caption{\scriptsize FSM for \roleFmt{a} \kern-6em}
    \label{fig:fms_T_1}
  \end{subfigure}
  \hspace{1.2em}
  \begin{subfigure}{.27\linewidth}
    \centering
    \begin{tikzpicture}[>=to, -to, thick, font=\tiny,
        initial left, initial distance=1.5ex, initial text={},
        state/.style={draw, circle, node distance=1cm and 1.6cm, inner sep=3pt},
        every path/.style={every node/.style={fill=white}}]

      \node[state,initial] (1) {};
      \node[state] (2) [right=of 1] {};
      \node[state] (3) [below=of 2] {};

      \path (1) edge node {$\stFmt{\roleFmt{q}?{\stLabFmt{upd}}_{\roleQ}}$} (2);
      \path (2) edge node {$\stFmt{\roleFmt{p}?{\stLabFmt{upd}}_{\roleP}}$} (3);
      \path (3) edge[bend right=15] node {$\stFmt{\roleFmt{m}!\stLabFmt{std}}$} (1);
      \path (3) edge[bend left=25] node {$\stFmt{\roleFmt{m}!\stLabFmt{wtd}}$} (1);

    \end{tikzpicture}
    \caption{\scriptsize Reordered FSM for \roleFmt{a} \kern-5em}
    \label{fig:fms_T_1_reordering}
  \end{subfigure}
  \hfill
  \begin{subfigure}{.33\linewidth}
    \centering
    \begin{tikzpicture}[>=to, -to, thick, font=\tiny,
        initial left, initial distance=1.5ex, initial text={},
        state/.style={draw, circle, node distance=1cm and 1.6cm, inner sep=3pt},
        every path/.style={every node/.style={fill=white}}]

      \node[state,initial] (1) {};
      \node[state] (2) [right=of 1] {};
      \node[state] (3) [below=of 2] {};

      \path (1) edge node {$\stFmt{\roleFmt{p}?{\stLabFmt{upd}}_{\roleP}}$} (2);
      \path (2) edge node {$\stFmt{\roleFmt{q}?{\stLabFmt{upd}}_{\roleQ}}$} (3);
      \path (3) edge node {$\stFmt{\roleFmt{m}!\stLabFmt{std}}$} (1);

    \end{tikzpicture}
    \caption{\scriptsize Covariant FSM for \roleFmt{a}\kern-2em}
    \label{fig:fms_T_1_branching}
    \label{fig:fms_T_1_covaraince}
  \end{subfigure}
  \caption{\small FSMs for role \roleFmt{a}}
  \label{fig:fms_1_2}
\end{figure}

\section{Motivation and Background}
\label{sec:ams}
\label{SEC:AMS}

{\bf Motivation. }  \Cref{fig:federated_protocol} depicts the message exchange coordinating distributed model training, referred to as the \emph{federated learning protocol}, reflecting communication patterns in centralised federated learning systems~\cite{DBLP:journals/ftml/KairouzMABBBBCC21}.  
The protocol involves a model store \roleFmt{m}, clients $\roleP[1], \ldots, \roleP[n]$, and an aggregator \roleFmt{a}, 
and proceeds in phases as follows: 
\begin{enumerate*}
\item \emph{Model Distribution}:  the model store sends the current global model to all clients; 
\item \emph{Update Submission}: each client performs local computation and sends an update to the aggregator; and 
\item \emph{Aggregation Result}: the aggregator combines the updates and sends either a standard or weighted result to the model store. 
\end{enumerate*}
The model store then updates the global model and redistributes it for the next round.

We consider a minimal instance of the federated learning protocol with two clients $\roleP$ and $\roleQ$, 
enforcing an ordered submission of updates to the aggregator $\roleFmt{a}$, where $\roleP$ precedes $\roleQ$. 

From the perspective of~\roleFmt{a} under \emph{synchronous} communication,  
the local protocol can be represented by a finite state machine (FSM),  as shown in~\Cref{fig:fms_T_1}, 
with the following steps:
\begin{enumerate*}
\item 
\label{eq:step_1} 
receive (\stFmt{?}) ${\stLabFmt{upd}}_{\roleP}$ from \roleP; 
\item receive ${\stLabFmt{upd}}_{\roleQ}$ from \roleQ; 
\item send (\stFmt{!}) aggregated result -- either \stLabFmt{std} or \stLabFmt{wtd} --  to the model store \roleFmt{m}; and 
\item  repeat from step~\ref{eq:step_1}. 
\end{enumerate*}

In practice, communication is  \emph{asynchronous} and typically modelled as FIFO message passing. 
 In this scenario, $\roleFmt{a}$ may receive the update from $\roleQ$ before that from $\roleP$, violating the ordering in~\Cref{fig:fms_T_1} and necessitating \emph{refinement} to account for \emph{message reordering}. 
 The refined FSM, shown in~\Cref{fig:fms_T_1_reordering}, 
 captures this by admitting the reception order $\roleQ$ followed by $\roleP$, while preserving the subsequent aggregation and response behaviour. 
 As the updates from $\roleP$ and $\roleQ$ are independent, the two FSMs are refinements of each other.

\begin{figure*}[!t]
 \centering
\includegraphics[width=1.1\textwidth,height=0.26\textheight,keepaspectratio]{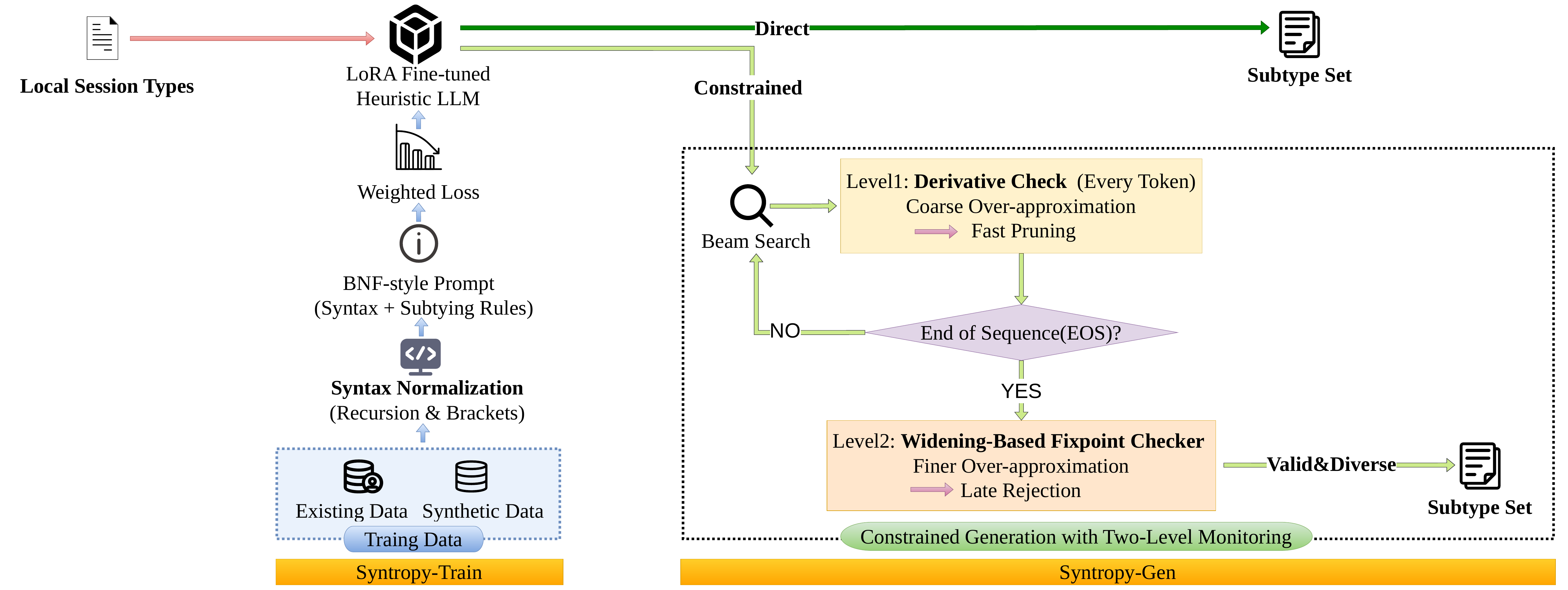}
\caption{Overview of \theTool}
\label{fig:overview_of_framework}
\end{figure*}

Furthermore, additional refinements are possible. The general protocol in~\Cref{fig:federated_protocol} can be viewed as a \emph{contravariant} refinement of the minimal instance, as it admits a broader class of inputs, while \emph{covariant} refinement arises when the aggregator deterministically produces a single fixed result~(\Cref{fig:fms_T_1_covaraince}). 

Such refinements occur naturally in distributed software systems, where communication patterns are commonly adapted for efficiency, \eg a service may omit certain responses to reduce latency. 
Nevertheless, even minor changes may introduce severe communication inconsistencies. For example, refining the behaviour of $\roleFmt{a}$ to exclude updates from $\roleQ$, while leaving other participants unaffected, yields a communication system that violates compatibility, leading to deadlock. 
This highlights the challenge of designing correct refinements and motivates the application of formal specifications to ensure the rigorous synthesis of protocol refinements that preserve communication properties.

\vspace{.3em}
\noindent
{\bf \MPST and Asynchronous Subtyping. } Multiparty Session Types (\MPST)~\cite{HYC16,POPL19LessIsMore} provide a framework for specifying and verifying communication protocols. 
In \MPST, the communication behaviour of each role (denoted by $\roleP, \roleQ, \roleFmt{s}, \rolePi, \ldots \in \roleSet$) is described by 
\emph{local session types} (or simply \emph{session types}), and protocol refinement is formalised as a subtyping relation on these types. 
In particular, \emph{asynchronous multiparty subtyping} (AMS)~\cite{GPPSY2023} offers a sound and complete declarative characterisation of this relation, ensuring  that subtypes -- \ie type-based protocol refinements --  can safely replace their supertypes while preserving all desirable communication safety properties. 

The syntax of session types, ranging over $\stT, \stTi, \stT[i], \ldots$, is inductively defined as follows: 

\vspace{.2em}
\centerline{\(
\stT \bnfdef  \stFmt{\oplus_{i \in I}\, \roleP ! \stChoice{\stLab[i]}{} \stSeq \stT[i]} 
        \ \bnfsep \stFmt{{\&}_{i \in I}\, \roleP ? \stChoice{\stLab[i]}{} \stSeq \stT[i]}  \ \bnfsep  \stRec{\stRecVar}{\stT} \ \bnfsep \ \stRecVar \ 
           \bnfsep  \stEnd
\)}

\vspace{.1em}
\noindent
The constructs $\stFmt{\oplus_{i \in I}\, \roleP ! \stChoice{\stLab[i]}{} \stSeq \stT[i]}$ and  
$\stFmt{{\&}_{i \in I}\, \roleP ? \stChoice{\stLab[i]}{} \stSeq \stT[i]}$ denote \emph{internal} and \emph{external choices}, corresponding to sending ($\stFmt{!}$) to or receiving ($\stFmt{?}$) from  role $\roleP$ a label 
$\stLab[i]$,  respectively, where $I \neq \emptyset$ and the labels $\stLab[i]$ are pairwise distinct. The type 
$\stEnd$ marks \emph{termination} (omitted when unambiguous), while $\stRec{\stRecVar}{\stT}$ introduces recursion  with variable $\stRecVar$. 

The behaviour of the aggregator $\roleFmt{a}$ in the minimal federated learning protocol is captured by 

\vspace{.25em}
\centerline{\(
\stT[\roleFmt{a}] = \stRec{\stRecVar}{\stFmt{\roleP ? {\stLabFmt{upd}}_{\roleP} \stSeq \roleFmt{q} ? {\stLabFmt{upd}}_{\roleQ} \stSeq 
\oplus\{ \roleFmt{m} ! \stLabFmt{std} \stSeq \stRecVar, \roleFmt{m} ! \stLabFmt{wtd} \stSeq \stRecVar\}}}, 
\)}

\vspace{.25em}
\noindent
which is an alternative representation of the FSM in~\Cref{fig:fms_T_1}.

In AMS, besides covariance (fewer output options) ({\bf C1}) and contravariance (more input options) ({\bf C2}), two forms of asynchronous message reordering are  allowed, whereby a subtype may anticipate input and output actions of the supertype: 
\begin{itemize}[leftmargin=-1pt, nosep]
\item[] {\bf R1.} 
An input from $\roleP$ may be anticipated before a finite number of inputs not from $\roleP$; 
\item[] {\bf R2.} 
An output to $\roleP$ may be anticipated before a finite number of inputs (from any participant), and before outputs not directed to $\roleP$. 
\end{itemize}

To formalise AMS, session types are interpreted as (possibly infinite) session trees. 
For instance, Fig.~\ref{fig:session_tree}\subref{fig:session_tree_T1} presents the session tree $\stFmt{{\mathbb{T}}_{\roleFmt{a}}}$ 
 associated with $\stT[\roleFmt{a}]$. 
A coinductive tree refinement relation $\stRefinement$ is defined on such session trees, 
and the asynchronous subtyping relation $\stSub$ is obtained by lifting 
$\stRefinement$ to session types. 

Consider the session type (equivalent to FSM in~\Cref{fig:fms_T_1_reordering}) 
 
 \vspace{.25em}
 \centerline{\(
\stTi[\roleFmt{a}] = \stRec{\stRecVar}{\stFmt{\roleFmt{q} ? 
{\stLabFmt{upd}}_{\roleQ} \stSeq \roleFmt{p} ? {\stLabFmt{upd}}_{\roleP} \stSeq 
\oplus\{ \roleFmt{m} ! \stLabFmt{std} \stSeq \stRecVar, \roleFmt{m} ! \stLabFmt{wtd} \stSeq \stRecVar\}}}, 
\)}

\vspace{.25em}
\noindent
obtained from $\stT[\roleFmt{a}]$ by reordering the inputs. 
Its session tree $\stFmt{{\mathbb{T}'}_{\!\!\roleFmt{a}}}$, shown in Fig.~\ref{fig:session_tree}\subref{fig:session_tree_T_2}, 
is related to $\stFmt{{\mathbb{T}}_{\roleFmt{a}}}$ by $\stRefinement$; 
 hence $\stTi[\roleFmt{a}] \stSub \stT[\roleFmt{a}]$, indicating that  $\stTi[\roleFmt{a}]$ can safely substitute for $\stT[\roleFmt{a}]$. 

AMS is undecidable in general~\cite{DBLP:journals/iandc/BravettiCZ17,DBLP:conf/fossacs/LangeY17}. 
Together with its intrinsic complexity, this renders the synthesis of asynchronous subtypes substantially non-trivial and inherently error-prone, giving rise to the main research question of this paper:

\vspace{-0em}
\begin{tcolorbox}[colback=gray!6, colframe=gray]
{{\emph{How can subtypes be synthesised automatically under asynchronous multiparty subtyping?}}}
\end{tcolorbox}
\vspace{-0em}

\noindent
To tackle this challenge, we propose a system based on large language models, described in the next section.

\algrenewcommand\algorithmicrequire{\textbf{Input:}}
\algrenewcommand\algorithmicensure{\textbf{Output:}}
\lstset{
    basicstyle=\ttfamily\small,
    numbers=left,
    numberstyle=\tiny\color{gray},
    frame=single,
    mathescape=true,
    breaklines=false
}

\section{\theTool: A Framework for Asynchronous Subtype Synthesis}
\label{sec:framework} 
\label{SEC:FRAMEWORK}

Figure~\ref{fig:overview_of_framework} illustrates the overall architecture of \theTool, 
a framework for automatically synthesising asynchronous subtypes from a source session type. 
It consists of two complementary modules: \trainPhase, which trains large language models (LLMs) to learn subtype generation patterns,   
and \inferencePhase, which leverages the trained model to produce diverse asynchronous subtypes guided by the 
asynchronous multiparty subtyping introduced in~\Cref{sec:ams}, thereby improving syntactic and semantic consistency. 


\vspace{-0em}
\subsection{\trainPhase: Learning Asynchronous Subtype Generation}
\label{sec:framework_training_phase}
\label{sec:training_phase}

We fine-tune an open-source model (\eg Qwen2.5-Coder-7B-Instruct \cite{hui2024qwen25codertechnicalreport}) using LoRA to generate subtypes conditioned on a given session type (the \emph{supertype}). Since a supertype may correspond to an unbounded number of subtypes under asynchronous subtyping, the model is trained to capture structure-preserving transformations and generate diverse and structurally distinct, valid outputs.

Our training data consists of session type specifications derived from \MPST literature~\cite{DBLP:conf/concur/BravettiCLYZ19,DBLP:conf/tacas/BocchiKM24,DBLP:conf/ppopp/CutnerYV22,DBLP:conf/concur/BocchiK0025}, augmented with synthetically generated examples to improve coverage and diversity (see~\Cref{sec:dataset} for details).  
Each instance is of the form $\{\stS, [\{\stT[1], \trainLab[1]\}, \ldots, \{\stT[n], \trainLab[n]\}], n\}$, 
where $\stS$ denotes the supertype, $n$ the number of subtypes, and for each $i \leq n$, $\stT[i]$ is a subtype with auxiliary (heuristic) label  $\trainLab[i]$~(\eg indicating one or multiple transformations). To accommodate variable-length instances, 
 we employ dynamic padding during training for efficient batching with minimal overhead.  This design exposes the model to diverse transformation patterns, allowing it to generate valid subtypes beyond a fixed set of explicitly defined rules. 

To facilitate training, we adapt the syntax of session types in~\Cref{sec:ams} into a model-friendly representation:

\vspace{.25em}
\centerline{\(
\begin{array}{r@{\quad}c@{\quad}l@{\quad}l}
\stT &\bnfdef&  \stFmt{\roleP ! \stLab; \stT}  \bnfsep \stFmt{\roleP ? \stLab; \stT}  \bnfsep \stEnd&
 \\[.1em]
& \bnfsep & \stFmt{\mpFmt{\texttt{lbrace}} \;\roleP ! \stLab[1]; \stT[1], \ldots, \roleP ! \stLab[n]; \stT[n]\; \mpFmt{\texttt{rbrace}}} &
\\[.1em]
&  \bnfsep & \stFmt{\mpFmt{\texttt{lbrace}} \;\roleP ? \stLab[1]; \stT[1], \ldots, \roleP ? \stLab[n]; \stT[n]\; \mpFmt{\texttt{rbrace}}} & 
\\[.1em]
&\bnfsep & \stFmt{\mpFmt{\texttt{REC}\_\stFmt{\mathbf{X}}\_\texttt{OPEN}} \;\stT} \bnfsep \stFmt{\mpFmt{\texttt{REC}\_\stFmt{\mathbf{X}}\_\texttt{CLOSE}}}  & 
\end{array}
\)}

\vspace{.2em}
\noindent
Send and receive actions are explicitly represented in sequential form as 
$\stFmt{\roleP ! \stLab; \stT}$ and $\stFmt{\roleP ? \stLab; \stT}$, 
yielding a well-defined action structure suitable for sequence modelling; 
internal and external choices use $\texttt{lbrace} \;\stFmt{\cdot}\; \texttt{rbrace}$;   
recursion is expressed by 
$\stFmt{\mpFmt{\texttt{REC}\_\stFmt{\mathbf{X}}\_\texttt{OPEN}} \;\stT}$, 
binding $\stFmt{\mathbf{X}}$ over $\stT$, with recursive occurrences encoded as 
$\stFmt{\mpFmt{\texttt{REC}\_\stFmt{\mathbf{X}}\_\texttt{CLOSE}}}$, 
each denoting a continuation to the corresponding binder. 
These modifications preserve semantics while making structural boundaries explicit and reducing syntactic ambiguity.

\begin{table}[t]
\footnotesize
\centering
\caption{Transformations encoded in prompts}
\label{tab:prompt}
\vspace{-1em}
\renewcommand{\arraystretch}{1.2}

\begin{tabular}{p{0.1\linewidth} p{0.25\linewidth} p{0.49\linewidth}}
\toprule
\textbf{Rule} & \textbf{Prompt} & \textbf{Example} \\
\midrule

\Identity &
Subtype \textcolor{purple!60}{identical} to  supertype. &
-- \\

\midrule

\Unfold &
Unfold recursion via \textcolor{purple!60}{substitution} of the  \newline recursive variable.  &
\begin{minipage}[t]{\linewidth}
\textsc{Valid}
\\[.3em]
$\begin{aligned}
& {\scriptsize \stFmt{\mpFmt{\texttt{REC}\_\stFmt{\mathbf{X}}\_\texttt{OPEN}} \;\roleP ? \stLab; \roleP ! \stLabi; \mpFmt{\texttt{REC}\_\stFmt{\mathbf{X}}\_\texttt{CLOSE}}}
\Rightarrow }
\\[-.5em]
& {\scriptsize \stFmt{\roleP ? \stLab; \roleP ! \stLabi;  \mpFmt{\texttt{REC}\_\stFmt{\mathbf{X}}\_\texttt{OPEN}} \;\roleP ? \stLab; \roleP ! \stLabi;   \mpFmt{\texttt{REC}\_\stFmt{\mathbf{X}}\_\texttt{CLOSE}}}}
\end{aligned}$
\end{minipage}
 \\

\midrule
\begin{minipage}[t]{\linewidth}
\RefA 
\end{minipage}
&
\begin{minipage}[t]{\linewidth}
Move recv $\stFmt{\roleP?\stLab}$ \textcolor{purple!60}{earlier} when roles differ.
\end{minipage}
&
\begin{minipage}[t]{\linewidth}
\textsc{Valid} \quad $\stFmt{\roleP?\stLab; \roleQ?\stLabi; \stT \stSub \roleQ?\stLabi; \roleP?\stLab; \stT}$ \\[0.2em]
\textcolor{gray}{\textsc{Invalid}} \,$\stFmt{\roleP?\stLab; \roleP?\stLabi; \stT \stSub \roleP?\stLabi; \roleP?\stLab; \stT}$
\end{minipage}
\\
\midrule
\begin{minipage}[t]{\linewidth}
\RefB 
\end{minipage}
&
\begin{minipage}[t]{\linewidth}
Move send $\stFmt{\roleP!\stLab}$ \textcolor{purple!60}{earlier} across independent \newline actions. 
\end{minipage}
&
\begin{minipage}[t]{\linewidth}
\textsc{Valid} \quad $\stFmt{\roleP!\stLab; \roleQ?\stLabi; \stT \stSub \roleQ?\stLabi; \roleP!\stLab; \stT}$ \\[0.2em]
\textcolor{gray}{\textsc{Invalid}} \,$\stFmt{\roleP!\stLab; \roleP!\stLabi; \stT \stSub \roleP!\stLabi; \roleP!\stLab; \stT}$
\end{minipage}
\\
\midrule
\RefOut &
Internal choice: \newline subtype offers \textcolor{purple!60}{fewer} labels. &
\begin{minipage}[t]{\linewidth}
\textsc{Valid}
\\[.3em]
$\begin{aligned}
& \stFmt{\mpFmt{\texttt{lbrace}} \;\roleP! \stLab; \stT[1]\; \mpFmt{\texttt{rbrace}}} \,\stSub
\\[-.5em]
&  \stFmt{\mpFmt{\texttt{lbrace}} \;\roleP! \stLab; \stT[1], \roleP! \stLabi; \stT[2]\;\mpFmt{\texttt{rbrace}}} 
\end{aligned}$

\vspace{0.2em}

\textcolor{gray}{\textsc{Invalid}}
\\[0.3em]
$\begin{aligned}
&\stFmt{\mpFmt{\texttt{lbrace}} \;\roleP! \stLab; \stT[1], \roleP! \stLabi; \stT[3]\; \mpFmt{\texttt{rbrace}}} \, \stSub
\\[-.5em]
& \stFmt{\mpFmt{\texttt{lbrace}} \;\roleP! \stLab; \stT[1], \roleP! \stLabi; \stT[2]\; \mpFmt{\texttt{rbrace}}}
\end{aligned}$
\end{minipage}
 \\
\midrule
\RefIn &
External choice: \newline subtype accepts \textcolor{purple!60}{more} labels. 
&
\begin{minipage}[t]{\linewidth}
\textsc{Valid}
\\[.3em]
$\begin{aligned}
&\stFmt{\mpFmt{\texttt{lbrace}} \;\roleP? \stLab[1]; \stT[1], \roleP? \stLab[2]; \stT[2], \roleP? \stLab[3]; \stT[3]\; \mpFmt{\texttt{rbrace}}} \stSub 
\\[-.5em]
& \stFmt{\mpFmt{\texttt{lbrace}} \;\roleP? \stLab[1]; \stT[1], \roleP? \stLab[2]; \stT[2]\; \mpFmt{\texttt{rbrace}}}
\end{aligned}$

\vspace{0.2em}
\textcolor{gray}{\textsc{Invalid}}
\\[0.3em]
$\begin{aligned}
&\stFmt{\mpFmt{\texttt{lbrace}} \;\roleP? \stLab[1]; \stT[1] \; \mpFmt{\texttt{rbrace}}} \stSub 
\\[-.5em]
& \stFmt{\mpFmt{\texttt{lbrace}} \;\roleP? \stLab[1]; \stT[1], \roleP? \stLab[2]; \stT[2]\; \mpFmt{\texttt{rbrace}}}
\end{aligned}$
\end{minipage}
\\
\bottomrule
\end{tabular}
\end{table}

For each training instance, we construct a prompt consisting of 
\begin{enumerate*}
\item a theoretical context, including transformation descriptions (\eg \Identity, \RefA, \RefB, \RefIn, \RefOut, and \Unfold), and
\item the instance-specific input. 
\end{enumerate*} 
The theoretical context is derived from the prompt components summarised in~\Cref{tab:prompt} and provides structured guidance for subtype generation. 
 \RefA and \RefB correspond to the message reordering rules {\bf R1} and {\bf R2} in~\Cref{sec:ams}, respectively, 
while \RefOut and \RefIn capture covariance ({\bf C1}) and contravariance ({\bf C2}). 
Additionally, \Identity instantiates the reflexivity of subtyping, ensuring the existence of at least one valid subtype for every supertype, 
namely itself, and \Unfold expands recursive types by recursively substituting bound variables.

\begin{example}[Multiple Transformations]
\label{ex:transformation}
Consider the supertype: 

\centerline{\(
\stS = \stFmt{\roleP ? \stLab[1]; \mpFmt{\texttt{lbrace}} \;\roleP ! \stLab[2]; \roleQ ? \stLab[3]; 
\roleR ? \stLab[4]; \stEnd, \roleP ! \stLab[5]; \stEnd\; \mpFmt{\texttt{rbrace}}}
\)}

\noindent 
Under \RefA, the subtype 

\centerline{\(
\stT[1] = \stFmt{\roleP ? \stLab[1]; \mpFmt{\texttt{lbrace}} \;\roleP ! \stLab[2]; \roleR ? \stLab[4]; \roleQ ? \stLab[3]; 
 \stEnd, \roleP ! \stLab[5]; \stEnd\; \mpFmt{\texttt{rbrace}}}
\)}

\noindent
is obtained, while applying \RefOut yields $\stT[2] = \stFmt{\roleP ? \stLab[1]; \roleP ! \stLab[5]}$. 

Note that $\stT[2]$ admits multiple transformations (\eg \RefA followed by \RefOut or directly via \RefOut), whereas the generator may assign only one as its heuristic label.
\end{example}



Training sequences follow a standard format consisting of a system prompt, a user prompt, and the corresponding assistant output. Prompt tokens are masked in the loss computation (\ie excluded from the objective), and loss is computed exclusively over the assistant outputs, corresponding to the generated subtype sequences. This ensures that the theoretical context serves as conditioning information rather than a target for imitation. 

We adopt a weighted token-level loss, assigning a weight of $1.0$ to subtype tokens and $0.2$ to auxiliary fields,  including  labels and the number of subtypes. Since both may originate from heterogeneous sources (\eg benchmark data or synthetic generation) and do not necessarily reflect canonical transformation patterns~(see~\Cref{ex:transformation}), they are treated as weak supervision signals. This encourages the model to prioritise learning structurally valid subtype sequences while mitigating overfitting to auxiliary information. 

Finally, the model is fine-tuned using LoRA (rank 64, $\alpha = 32$, dropout 0.05) with a maximum sequence length of 8k tokens. Training is performed using AdamW with a cosine learning rate schedule and a warmup ratio of 0.1 for 3 epochs.

\subsection{\inferencePhase: Synthesising Asynchronous Subtypes}
\label{sec:inference_phase}

\inferencePhase generates asynchronous subtypes of a given supertype using the trained model. 
It supports two generation strategies:  \emph{direct generation}, where the LLM produces subtype candidates guided by transformation rules adopted during training (\Cref{sec:training_phase}), and \emph{constrained generation}, 
which introduces two-level monitoring mechanisms to enforce asynchronous subtyping constraints and improve generation accuracy.

\vspace{-.5em}
\subsubsection{Direct Generation}
\label{sec:direct_generation}

In the direct generation approach, the trained model is prompted to synthesise subtypes. Generation is guided by transformation patterns -- namely \Identity, \RefA, \RefB, \RefIn, \RefOut, and \Unfold -- described in~\Cref{sec:training_phase}. These encourage the model to produce diverse candidates reflecting covariance, contravariance, and reordering.

In this process, subtypes are derived from the supertype through sequences of transformations, where successive applications expand the set of generated subtypes. Structured guidance is therefore provided for the LLM to generate pattern-compatible subtypes; however, semantic correctness is not guaranteed as this na\"ive approach relies entirely on the trained model.

\vspace{-.5em}
\subsubsection{Constrained Generation with Two-Level Monitoring}
\label{sec:constraint_generation}
To improve the semantic validity of generated subtypes while preserving diversity, 
we employ constrained generation with two-level monitoring. 
Our approach is inspired by the asynchronous multiparty subtyping framework of~\cite{DBLP:conf/concur/BocchiK0025}, which we adapt to guide both prefix-level filtering and final subtype verification within the constrained generation process.

The input supertype is first parsed into a session tree, a tree-structured representation of a session type, as illustrated in~\Cref{sec:ams}. 
Beam search then explores candidate subtypes, while the monitoring stages enforce semantic constraints throughout  generation.

We formalise the constrained generation procedure in~\Cref{alg:constrained_generation} and describe its main components below.

\begin{algorithm}[!t]
\caption{Constrained Generation with Two-Level Monitoring}
\label{alg:constrained_generation}
\begin{algorithmic}[1]
\small
\Require supertype $\stS$, language model $M$, beam size $k$
\Ensure set of candidate subtypes $\Omega$

\State $\stFmt{{\mathbb{T}}_{\stS}} \gets \textsc{Parse}(\stS)$
\State $\emph{\text{Beams}} \gets \{(\epsilon, 0.0)\}$
\State $C \gets \emptyset$

\For{each decoding step}
    \For{each $(\emph{\text{seq}}, \emph{\text{score}}) \in \emph{\text{Beams}}$}
        \For{each $(t,p) \in {\emph{\text{Top}}}_{2k}(M)$}        
        
            \State $\emph{\text{seq}}' \gets \emph{\text{seq}} \cdot t$
            \State $\stFmt{\mathbb{T'}} \gets \textsc{ParsePrefix}(\emph{\text{seq}}')$
            
           \Comment{Level~1 (Derivative Check)} \kern12em
            \If{$\stFmt{\mathbb{T'}} \neq \bot$ \textbf{and} \textsc{Feasible}($\stFmt{\mathbb{T'}}, \stFmt{{\mathbb{T}}_{\stS}}$)}
                \If{$t = \textsc{EOS}$}
                    \State $\stFmt{{\mathbb{T}}_{\text{f}}} \gets \textsc{Parse}(\emph{\text{seq}}')$
                    
                    \Comment{{Level~2 (Widening-Based Fixpoint Check)}} \kern2.8em
                    \If{$\textsc{SimCheck}(\stFmt{{\mathbb{T}}_{\text{f}}}, \stFmt{{\mathbb{T}}_{\stS}})$}
                        \State $\Omega \gets \Omega \cup \{\emph{\text{seq}}'\}$
                    \EndIf
                \Else
                    \State $C \gets C \cup \{(\emph{\text{seq}}', score + \log p)\}$
                \EndIf
            \EndIf
        \EndFor
    \EndFor
    \State $\emph{\text{Beams}}  \gets {\emph{\text{Top}}}_{k}(C)$
\EndFor
\end{algorithmic}
\end{algorithm}

\vspace{.3em}
\noindent
{\bf Level 1: Token-level Derivative Check. } 
Level 1 takes as input a decoded prefix together with the session tree $\stFmt{{\mathbb{T}}_{\stS}}$ of the 
input supertype $\stS$,  
and returns a Boolean indicating whether the prefix can still be extended to a valid subtype of $\stS$.

At each decoding step, the prefix is incrementally 
parsed into a partially constructed session tree $\stFmt{\mathbb{T'}}$, 
whose unresolved positions are represented by \texttt{Hole} nodes. 
Since the \MPST grammar is deterministic, each syntactically valid prefix corresponds to a \emph{unique} partial tree, while 
\texttt{Hole} nodes denote subtrees to be completed. The monitor then performs a lightweight coinductive derivative-based compatibility check between 
$\stFmt{{\mathbb{T}}_{\stS}}$ and $\stFmt{\mathbb{T'}}$ to 
determine whether such an extension remains possible. 
Specifically, it evaluates  a predicate 
$\textsc{Feasible}(\stFmt{\mathbb{T'}}, \stFmt{{\mathbb{T}}_{\stS}})$,  
which returns \textsc{true} if $\stFmt{\mathbb{T'}}$  admits at least one completion 
whose corresponding session type is a valid subtype of $\stS$, 
and \textsc{false} otherwise.

The feasibility check is based on derivative reasoning 
over session trees~\cite{DBLP:conf/concur/BocchiK0025}, applied to partial constructions.  
 It establishes whether the behaviour induced  by $\stFmt{\mathbb{T'}}$
is compatible with that represented by $\stFmt{{\mathbb{T}}_{\stS}}$. 
For \texttt{send} nodes, each branch generated in $\stFmt{\mathbb{T'}}$ 
must be supported by a corresponding transition in $\stFmt{{\mathbb{T}}_{\stS}}$;  
for \texttt{receive} nodes, realisability is preserved provided 
that at least one branch of $\stFmt{\mathbb{T'}}$ is admitted by $\stFmt{{\mathbb{T}}_{\stS}}$, 
possibly after unfolding recursion in $\stFmt{{\mathbb{T}}_{\stS}}$ to expose further input actions. 
A prefix is rejected only when no completion of the current partial tree can satisfy the subtyping constraints 
(\eg because the matched node of $\stFmt{{\mathbb{T}}_{\stS}}$ is \texttt{end}, \ie the terminal node).

This monitor yields a prefix-level \emph{over-approximation}: 
any explored prefix that could lead to a valid subtype is preserved, while some infeasible ones 
may also be retained. This property is relative to beam search, which does not exhaustively enumerate all paths, and to the generally infinite space of subtypes.  
 In practice, Level 1 serves as an efficient per-token filter within beam search. When 
$\textsc{Feasible}(\stFmt{\mathbb{T'}}, \stFmt{{\mathbb{T}}_{\stS}})$ 
evaluates to \textsc{false}, the corresponding beam is pruned immediately, eliminating infeasible regions of the search space while preserving all potentially valid candidates.

Candidates that reach an end-of-sequence (EOS) token while satisfying the Level 1 condition are forwarded to Level 2; those that produce EOS otherwise are discarded, while all others continue under Level 1 monitoring.


\vspace{.3em}
\noindent
{\bf Level 2: Widening-Based Fixpoint Checker. } 
When a candidate reaches an EOS token, 
it is parsed into a complete session tree $\stFmt{{\mathbb{T}}_{\text{f}}}$ 
and submitted to a full subtype checker against the input supertype  $\stS$. 
The checker, implemented as a function $\textsc{SimCheck}(\stFmt{{\mathbb{T}}_{\text{f}}}, \stFmt{{\mathbb{T}}_{\stS}})$, is built on a derivative-based decision procedure for asynchronous subtyping over session trees~\cite{DBLP:conf/concur/BocchiK0025}. 

The procedure explores pairs of states from $\stFmt{{\mathbb{T}}_{\text{f}}}$ and $\stFmt{{\mathbb{T}}_{\stS}}$, 
maintaining a worklist of obligations together with the information required to avoid unsound revisiting of recursive configurations. Each obligation is processed according to the structure of the current nodes. 
For \texttt{send} nodes, the action exposed by $\stFmt{{\mathbb{T}}_{\text{f}}}$  must be matched by a corresponding send derivative 
in $\stFmt{{\mathbb{T}}_{\stS}}$; for \texttt{receive} nodes, the checker requires that the branching structure of 
$\stFmt{{\mathbb{T}}_{\stS}}$ covers the inputs expected by $\stFmt{{\mathbb{T}}_{\text{f}}}$, 
in accordance with the asynchronous subtyping conditions. 
For terminal nodes, the check succeeds only if the corresponding configuration in $\stFmt{{\mathbb{T}}_{\stS}}$  
admits termination.

To ensure termination in the presence of recursion, 
the checker applies a \emph{widening operator} at designated recursion points, 
following the abstract interpretation framework of~\cite{DBLP:conf/concur/BocchiK0025}.  
Widening merges newly derived configurations with previously explored ones and detects recurring patterns, 
thereby ensuring convergence of the fixpoint computation. 
The procedure terminates when the worklist is exhausted. 
Acceptance then indicates that the session type represented by $\stFmt{{\mathbb{T}}_{\text{f}}}$
is a valid subtype of $\stS$.


This checker accepts only candidates satisfying the asynchronous subtyping relation, while some valid ones may be rejected.

\vspace{.3em}
\noindent
{\bf Correctness and Design Justification. } 
Each output subtype is required to pass both a coarse prefix-level filter (Level 1) and a finer-grained final verification step (Level 2). This two-level design confines computationally expensive subtype checking to complete candidates, while lightweight prefix-level filtering eliminates infeasible prefixes early during search. 

Level 2 operates as a conservative checker rather than a complete decision procedure, reflecting the undecidability of asynchronous subtyping: accepted candidates are valid, whereas rejection does not imply invalidity.


Additionally, this design promotes output diversity by enabling structurally distinct valid subtypes to emerge through different reorderings and interaction patterns permitted by asynchronous subtyping. An overly restrictive validation step would limit the search space explored by beam search and reduce the diversity of generated subtypes. Overall, the two-level design balances semantic correctness with adequate exploration of the candidate space.

\vspace{.3em}
\noindent
{\bf Implementation Details. }  We implement Algorithm~\ref{alg:constrained_generation} using beam search with beam size 
$k$. At each decoding step, each beam is expanded using the top-$2k$ tokens to compensate for pruning by the Level~1 filter and preserve diversity. Candidate sequences are ranked by cumulative log-probability, and the top-$k$ candidates are retained after each step. 

\section{Evaluation}
\label{sec:evaluation}

To comprehensively assess \theTool, we conduct a systematic empirical evaluation addressing the following research questions. 

\begin{itemize}[leftmargin=0pt, nosep]
\item[] {\bf RQ1. } How well does \theTool perform on existing benchmarks, and how does it generalise across different LLMs?
\vspace{.4em}
\item[] {\bf RQ2. } How does the performance of \theTool vary across LLMs fine-tuned with LoRA using training sets of different sizes? 
\vspace{.4em}
\item[] {\bf RQ3. }  How do the \trainPhase module, the BNF-style prompt, and the two-level monitoring mechanism in \inferencePhase  affect overall performance of \theTool?
\vspace{.4em}
\item[] {\bf RQ4. } How robust is \theTool under different subtyping transformations, including covariance and contravariance (collectively referred to as variance), as well as reordering?


\end{itemize}

\subsection{Experimental Setup}
\label{sec:experiment_setup}

\begin{figure*}[htbp]
    \centering
    \begin{minipage}[t]{0.45\textwidth}
        \vspace{0pt}
        \centering
        \includegraphics[width=1.0\linewidth]{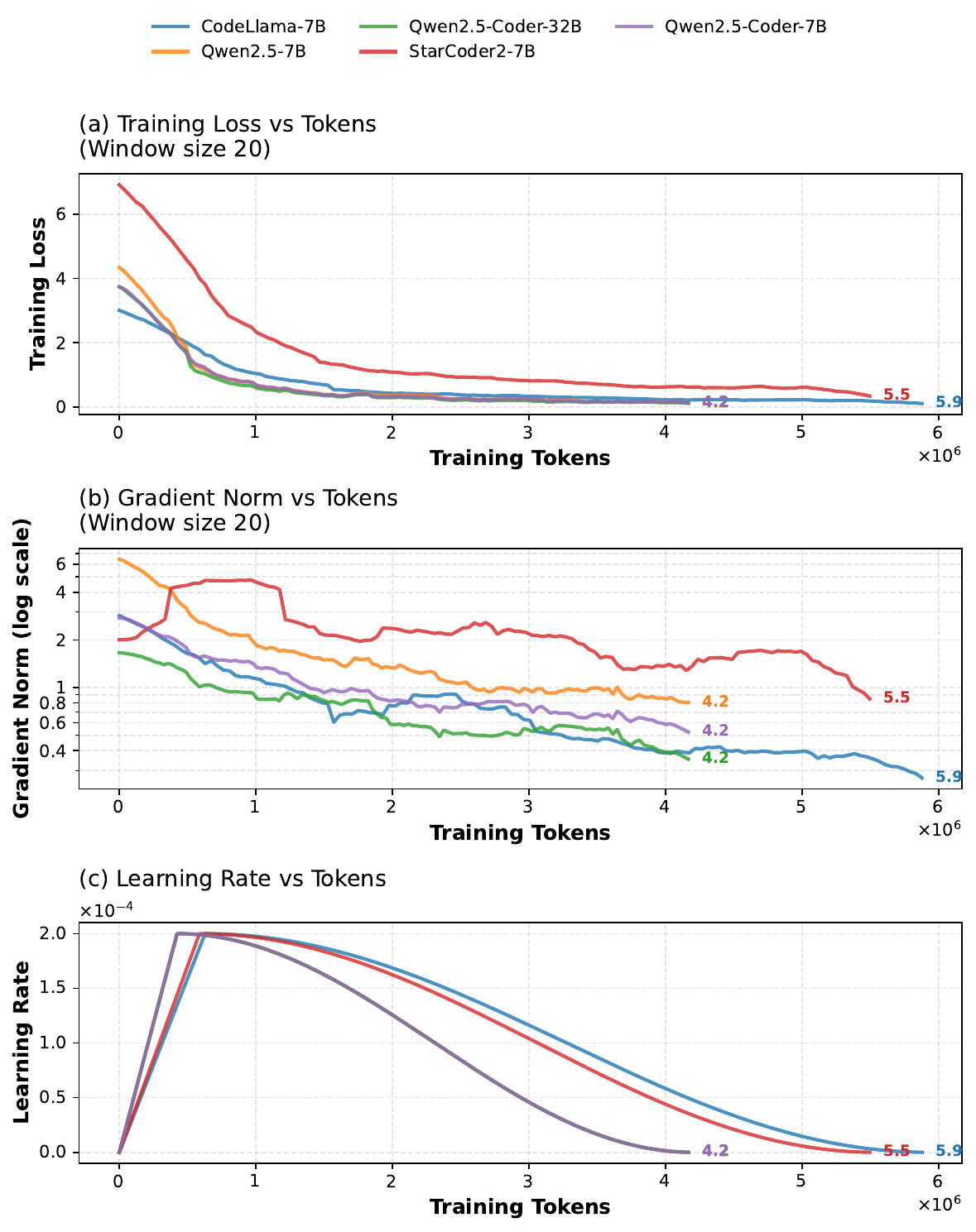}
        \vspace{-1.7em}
        \caption{\small Training dynamics across LLMs}
        \label{fig:training_llms}
    \end{minipage}
    \hfill
    \begin{minipage}[t]{0.45\textwidth}
        \vspace{0pt}
        \centering
        \includegraphics[width=1.0\linewidth]{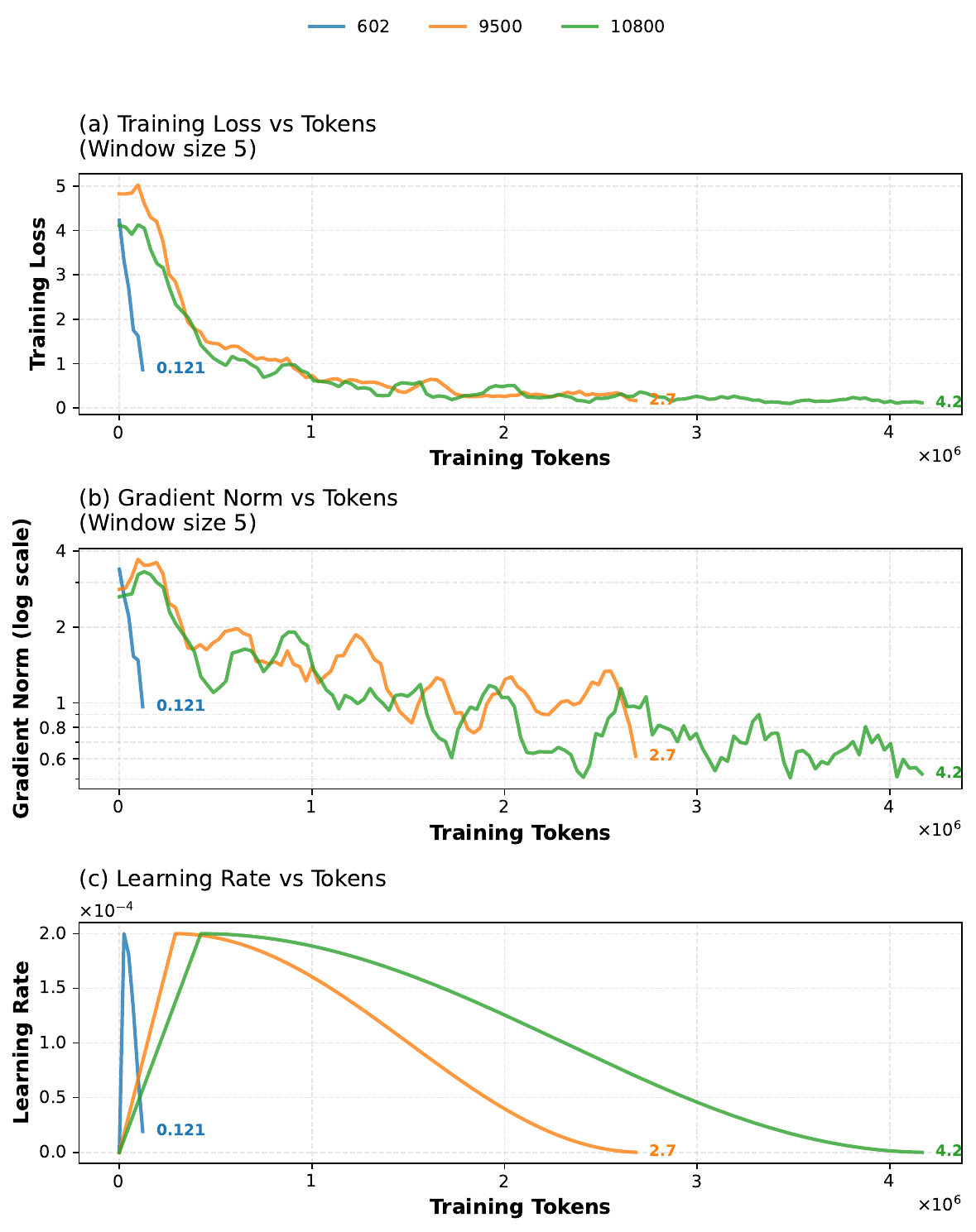}
         \vspace{-1.7em}
        \caption{\small Training dynamics across data scales} 
        \label{fig:training_data_scaling}
    \end{minipage}
\end{figure*}

\subsubsection{Dataset Construction}
\label{sec:dataset}
The dataset is constructed by collecting \emph{supertypes}, 
which serve as inputs to the framework. Two sources  are considered: \emph{literature-derived data} and \emph{synthetic data}.

The literature-derived data (\ie existing data) are collected from prior work on Multiparty Session Types (\MPST)~\cite{DBLP:conf/concur/BravettiCLYZ19,DBLP:conf/tacas/BocchiKM24,DBLP:conf/ppopp/CutnerYV22,DBLP:conf/concur/BocchiK0025}, covering real-world communication protocols in domains such as financial transactions, security, distributed coordination, and service orchestration. These protocols provide realistic and semantically meaningful specifications. 

To complement these examples and improve structural coverage, synthetic supertypes are additionally constructed to capture a broader range of communication patterns and structures not present in existing benchmarks, enabling evaluation across diverse protocols with varying levels of complexity.

\vspace{.3em}
\noindent
{\bf Subtype Generation and Validation. } 
To construct supervised training data, pairs of the form $(\text{supertype}, \text{subtype})$ are created, defining the learning task. 
For synthetic supertypes, subtypes are generated through a heuristic procedure inspired by 
an asynchronous subtyping algorithm~\cite{DBLP:conf/concur/BocchiK0025} and validated using its checker. 

 For literature-derived supertypes, no canonical target subtype is available.  Benchmark cases contain a limited number of reference subtypes, typically one per supertype, providing only a partial view of the refinement space.  To enrich the training data, additional subtypes are generated using the same procedure and validated accordingly. Validation results show that 99.13\% of subtypes from literature-derived supertypes (excluding original benchmark subtypes) and 97.97\% from synthetic supertypes are accepted, indicating high reliability.

Notably, the asynchronous subtyping algorithm in~\cite{DBLP:conf/concur/BocchiK0025} is sound but \emph{not} complete, and the checker inherits this limitation. Consequently, rejected subtypes may be deemed semantically valid. In addition, practical constraints prevent the checker from handling protocols beyond a certain size threshold (\eg 701 grammar units), such that even identity subtypes may be rejected.

\vspace{.3em}
\noindent
{\bf Dataset Split. } 
Supertypes are partitioned into training and test sets. The training set comprises 704 supertypes, yielding 10,800 $(\text{supertype}, \text{subtype})$ pairs (100 from literature-derived data and 604 from synthetic data). To balance learning signals, subtypes generated by different subtyping rules are approximately balanced. The test set consists of 100 representative supertypes (43 from literature-derived data and 57 from synthetic data) and is used as input for evaluation. 

\vspace{-.6em}
\subsubsection{Foundation Model Selection}
\label{sec:basic_model}
Qwen2.5-Coder-7B-Instruct~\cite{hui2024qwen25codertechnicalreport} serves as  the primary foundation model for our experiments.  
Unlike a raw pre-trained base model, it is instruction-tuned and optimised for code and structured generation, making it well suited for grammar-constrained protocol specifications such as \MPST session types.  The model balances generation quality and computational efficiency, enabling systematic fine-tuning and controlled ablation studies within reasonable resource constraints.   For reproducibility, a fixed random seed (seed = $42$) is used across all experiments. 

To the best of our knowledge, no prior work has explored the use of large language models for type-based generation in the context of \MPST. Existing research primarily focuses on formal methods rather than 
learning-based approaches. As a result, no directly comparable baseline method is available.

\vspace{-.5em}
\subsubsection{Evaluation Metrics}
\label{sec:overall_metrics}

To comprehensively assess the performance of \theTool, a multi-metric evaluation framework is adopted, targeting two main objectives: \emph{validity} and \emph{diversity}. Validity is evaluated at both syntactic and semantic levels, while diversity captures variations induced by subtyping transformations, including interaction reordering and variance. Semantic validity is further analysed by transformation type. Five metrics are applied: 

\begin{enumerate}[leftmargin=14pt, nosep]
\item {\bf Syntactic Validity.  } 
\label{item:metric_syn_validity}
The proportion of generated subtypes that conform to the formal grammar specification, as verified by a dedicated syntax checker. 

\item {\bf Semantic Validity. }  
\label{item:metric_sem_validity}
The proportion of syntactically valid subtypes accepted by the subtyping checker in~\cite{DBLP:conf/concur/BocchiK0025}, used as a proxy for semantic correctness. Since the checker is sound but not complete, this metric provides a lower bound.

\item {\bf Reordering vs. Variance Validity. } 
\label{item:metric_reorder_variance}
The acceptance rates of reordering-based and variance-based transformations under the subtyping checker.

\item {\bf Transformation Rule Distribution. } 
\label{item:metric_distribution}
The frequency of subtyping transformation rules applied during subtype generation, including \Identity, \RefA, \RefB, \RefIn, \RefOut, and \Unfold (see~\Cref{sec:framework}). A subtype with multiple transformations contributes to each corresponding rule count. 

\item {\bf Output Complexity. }  
\label{item:metric_output}
The average numbers of branches and message exchanges  per subtype, reflecting the structural diversity of the generated outputs. 
 \end{enumerate}
 
 \noindent
In addition,  the comparison with frontier language models (\Cref{sec:comparison_gpt_deepseek}), 
employs the {\bf \emph{coverage}} metric, defined as the proportion of supertypes for which at least one subtype is generated. 
This metric  measures a model's applicability  across diverse protocol specifications.

\begin{table}[t]
\centering
\caption{\small Training cost across LLMs}
\label{tab:training_llms_cost}
\vspace{-1em}
\footnotesize
\setlength{\tabcolsep}{20pt} 
\begin{tabular}{lcccccccc}
\toprule

{\bf Model}
& {\bf Time} (min) & {\bf Tok} (M) \\
\midrule

\rowcolor{gray!15} Qwen2.5-Coder-7B
 & 20.33  & 4.2  \\

CodeLlama-7B
 & 41.19  & 5.9  \\

StarCoder2-7B 
& 32.51  & 5.5 \\

Qwen2.5-7B
& 32.55  & 4.2  \\

Qwen2.5-Coder-32B
& 61.61  & 4.2  \\
\bottomrule
\end{tabular}
\vspace{-1.5em}
\end{table}

\vspace{-.2em}
\subsubsection{Training Configuration}
\label{sec:training_setting}
Alongside Qwen2.5-Coder-7B-Ins- \linebreak
truct, we fine-tune four additional open-source language models, 
CodeLlama-7B-Instruct~\cite{DBLP:journals/corr/abs-2308-12950}, 
Qwen2.5-7B-Instruct~\cite{qwen2025qwen25technicalreport}, 
StarCoder2-7B~\cite{DBLP:journals/tmlr/LiAZMKMMALCLZZW23}, and 
Qwen2.5-Coder-32B-Instruct \cite{hui2024qwen25codertechnicalreport}, using the same Low-Rank Adaptation (LoRA) configuration with a rank of $64$, a learning rate of $2 \times 10^{-4}$, and a batch size of $1$.  
All models are instruction-tuned, except for StarCoder2-7B, for which no Instruct variant is publicly available.  
For brevity, the ``-Instruct" suffix is omitted throughout the remainder of the paper, including figures and tables. 

Training is conducted for $3$ epochs, with costs reported in~\Cref{tab:training_llms_cost}.  
For Qwen2.5-Coder-7B,  the average sequence length is $2,027.5$ tokens per sample and the training throughput is approximately $3,415$ tokens per second. 

\Cref{fig:training_llms} illustrates the training loss curves for all models,  with steadily decreasing loss indicating stable optimisation dynamics. The gradient optimisation process remains stable, and the learning rate is well tuned, exhibiting near-optimal behaviour. 
These results  suggest that \theTool can be efficiently fine-tuned with moderate computational cost.

\subsection{RQ1. Benchmark Performance and Cross-LLM Generalisation}
\label{sec:evaluation_rq_1}

\begin{table*}[t]
\centering
 \caption{\small Performance and computational cost across LLMs}
\label{tab:overall_results}
\vspace{-1em}
\footnotesize
\setlength{\tabcolsep}{8pt}
\begin{tabular}{lcccccccc}
\toprule
& \multicolumn{4}{c}{{\bf No Monitoring}} & \multicolumn{4}{c}{{\bf With Monitoring}} \\
\cmidrule(lr){2-5} \cmidrule(lr){6-9}
{\bf Model}
& {\bf Syn.} (\%) & {\bf Sem.} (\%) & {\bf Br.} / {\bf Msg.} & {\bf Time} (min) / {\bf Tok} (M)
& {\bf Syn.} (\%) & {\bf Sem.} (\%) & {\bf Br.} / {\bf Msg.} & {\bf Time} (min) / {\bf Tok} (M) \\
\midrule

\rowcolor{gray!15}
Qwen2.5-Coder-7B
& 85.0 & 60.4 & 1.83 / 8.07 & 335.69 / 8.6
& 96.7 & 99.2 & 1.93 / 8.54 & 1041.56 / 27.5 \\

CodeLlama-7B
& 72.6 & 62.1 & 1.87 / 7.18 & 520.15 / 10.4
& 98.1 & 99.5 & 2.22 / 7.00 & 1733.71 / 33.5 \\

StarCoder2-7B
& 85.8 & 48.8 & 2.04 / 6.67 & 733.60 / 9.1
& 95.4 & 95.7 & 2.06 / 6.58 & 1212.12 / 30.2 \\

Qwen2.5-7B
& 80.4 & 60.9 & 1.90 / 7.32 & 585.19 / 8.6
& 97.0 & 96.4 & 2.01 / 7.22 & 1606.01 / 27.7 \\

Qwen2.5-Coder-32B
& 81.6 & 66.1 & 1.92 / 6.66 & 1940.00 / 8.5
& 96.3 & 95.6 & 1.99 / 7.23 & 3867.99 / 27.4 \\
\bottomrule
\end{tabular}
\end{table*}

\begin{figure*}[htbp]
    \centering
    \begin{minipage}[t]{0.33\textwidth}
        \vspace{0pt}
        \centering
        \includegraphics[width=\linewidth]{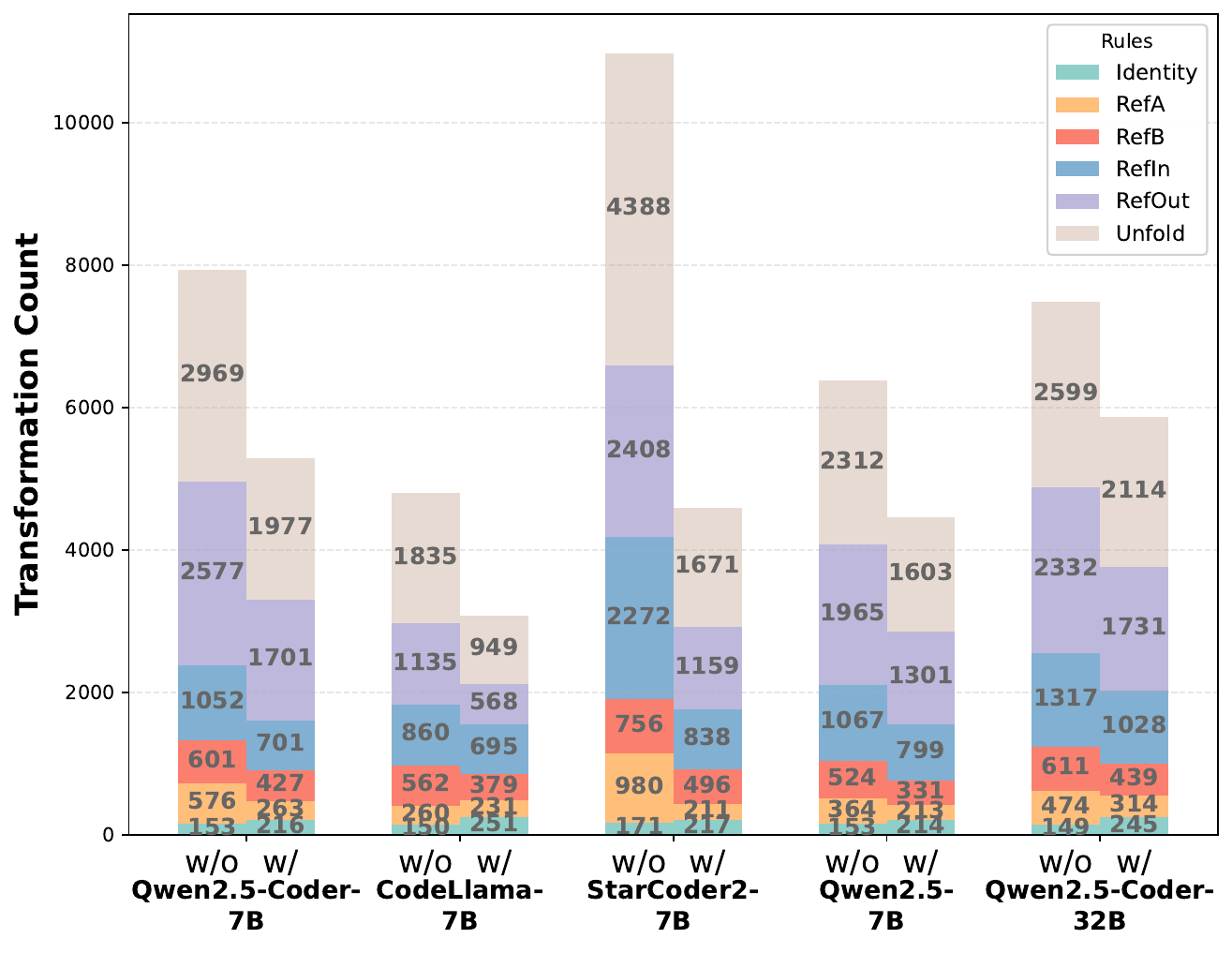}
        \caption{\small Transformation distribution across LLMs}
        \label{fig:distribution_llm}
    \end{minipage}
    \hfill
    \begin{minipage}[t]{0.65\textwidth}
        \vspace{0pt}
        \centering
        \begin{subfigure}[t]{0.51\linewidth}
            \centering
            \includegraphics[width=\linewidth]{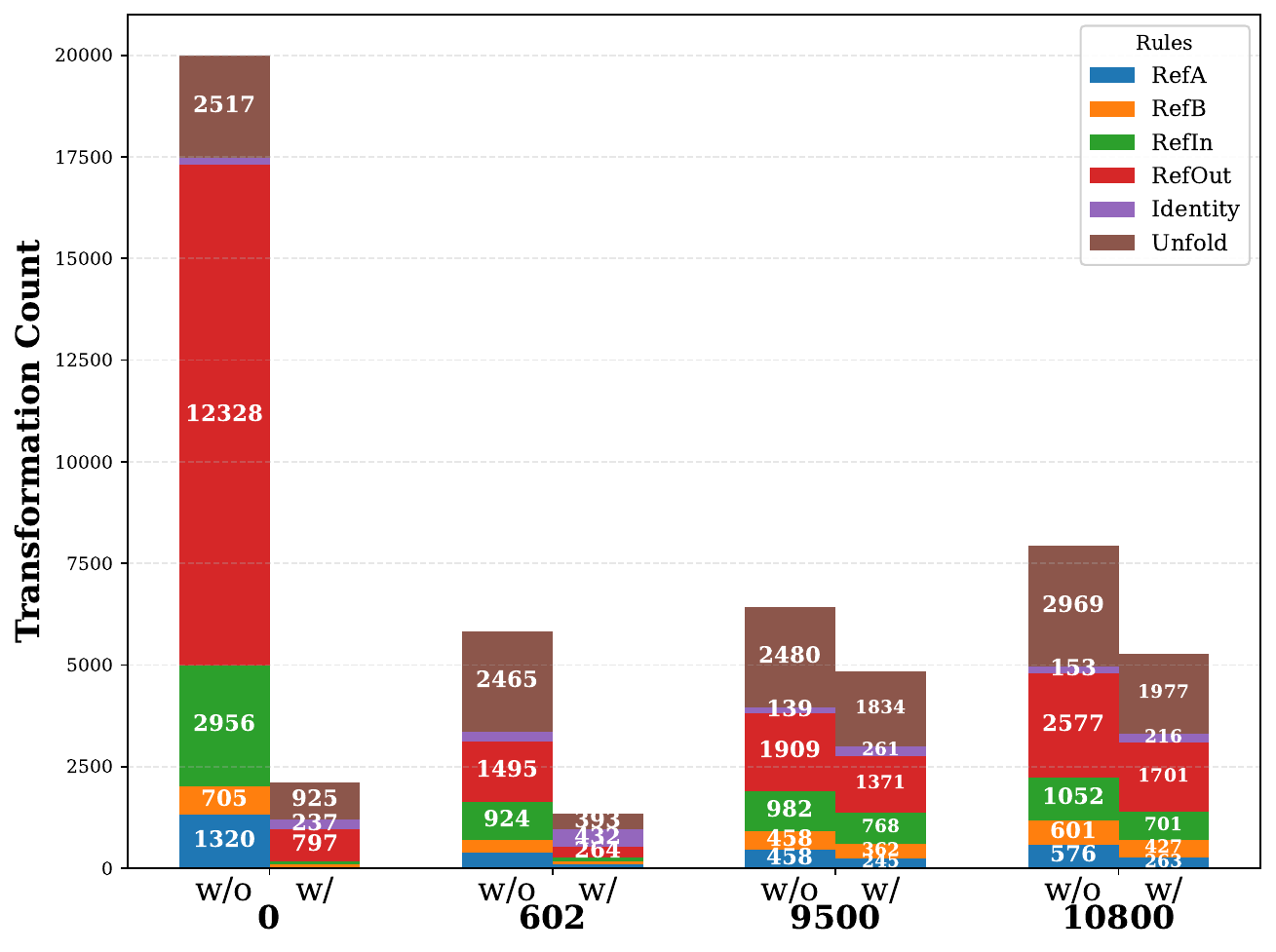}
            \caption{\footnotesize Transformation distribution across data scales}
            \label{fig:distribution_scaling}
        \end{subfigure}
        \hfill
        \begin{subfigure}[t]{0.47\linewidth}
            \centering
            \includegraphics[width=\linewidth]{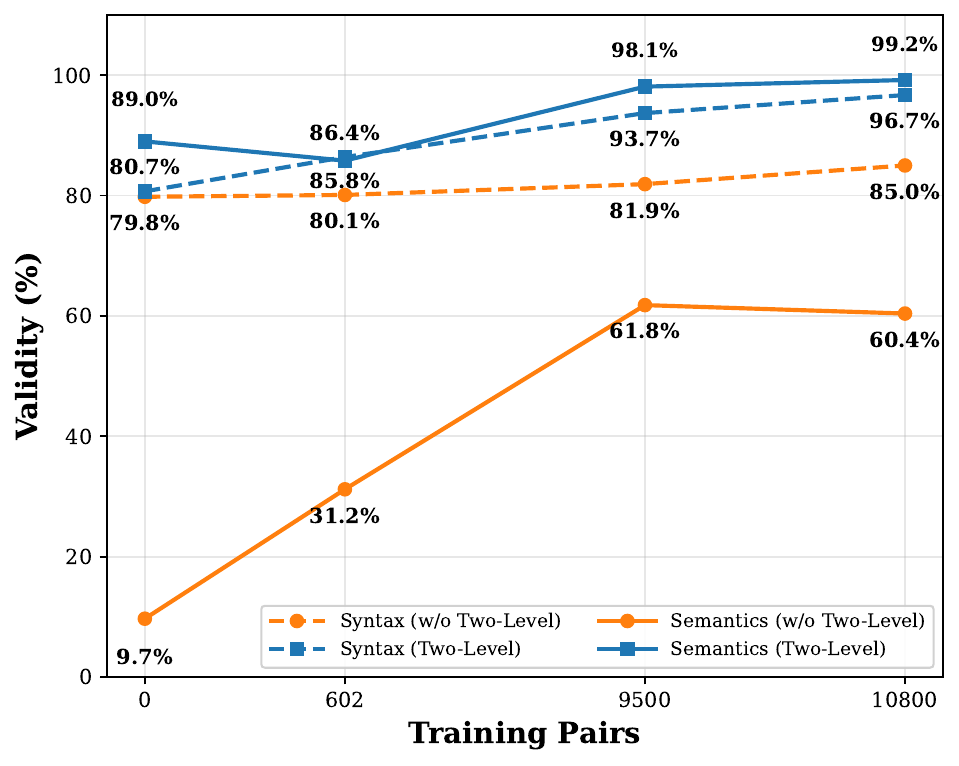}
            \caption{\footnotesize Validity across data scales}
            \label{fig:validity_scaling}
        \end{subfigure}
        \caption{\small Transformation distribution and validity across data scales}
        \label{fig:data_scaling_combined}
    \end{minipage}
\end{figure*}

We evaluate \theTool on 
our primary 
foundation model, 
Qwen2.5-Coder-7B,  using metrics defined  in~\Cref{sec:overall_metrics}, including syntactic and semantic validity, transformation rule distribution, and output complexity.  
We further assess its generalisation across additional LLMs, analysing performance consistency and cross-model applicability. 
\Cref{fig:distribution_llm} presents the transformation rule distribution, while the remaining metrics and computational cost are summarised in~\Cref{tab:overall_results}.


\vspace{-.5em}
\subsubsection{Results on the Primary Foundation Model}
\label{sec:result_Qwen2.5-Coder-7B}
We compare two approaches: \emph{direct generation} and \emph{constrained generation} (with Two-Level Monitoring), introduced in~\Cref{sec:inference_phase},  on the full test set. 

\vspace{.3em}
\noindent
{\bf Direct Generation (w/o Two-Level Monitoring). } 
 Syntactic validity reaches 85.0\%, while semantic validity  -- evaluated on syntactically valid outputs -- is 60.4\%, reflecting substantial semantic errors despite high syntactic correctness.
The transformation rule distribution indicates diverse application of subtyping rules, while each supertype yields approximately 5 subtypes per rule, with an average of 2 branching points and 8 message exchanges.
 
\vspace{.3em}
\noindent
{\bf Generation with Two-Level Monitoring. }  
 Syntactic validity increases to 96.7\%, while semantic validity among syntactically valid outputs reaches 99.2\%, indicating near-perfect semantic correctness. This improvement from 60.4\% to 99.2\% demonstrates the effectiveness of the monitoring-based approach, while the transformation rule distribution and output complexity remain comparable.

\vspace{.3em}
\noindent
{\bf Overhead Analysis. } 
We analyse the computational overhead introduced by Two-Level Monitoring. Enabling monitoring increases token consumption from 8.6M to 27.5M (approximately $3\times$), mainly due to additional validation and constraint enforcement.   
While runtime is reported in~\Cref{tab:overall_results}, it is not used as the primary cost metric due to its sensitivity to GPU allocation variability  in the HTC cluster; token-level overhead serves as the basis for subsequent analysis. 

Despite this additional cost, the monitoring mechanism substantially improves semantic validity, indicating a clear trade-off between efficiency and correctness. Direct generation is more suitable for efficiency-oriented scenarios, whereas monitoring-based generation is preferable for correctness-critical applications. 

\vspace{-.5em}
\subsubsection{Cross-LLM Generalisation}
\label{sec:model_generalisation}
We evaluate \theTool across multiple LLMs on the full test set, including CodeLlama-7B, \linebreak
StarCoder2-7B, Qwen2.5-7B, and Qwen2.5-Coder-32B. 
As shown in~\Cref{tab:overall_results}, syntactic validity remains consistently high (95.4\%--98.1\%), while semantic validity reaches up to 99.5\%, demonstrating robust generalisation across models. These results indicate that \theTool is not tied to a specific architecture and exhibits strong cross-model applicability.

The transformation rule distribution (\Cref{fig:distribution_llm}) is generally consistent, with some variations among models.  
In particular, CodeLlama-7B exhibits limited diversity under default decoding settings, failing to generate certain transformations (\eg \RefA and \RefB). To address this, we adopt stochastic beam search with temperature scaling ($T = 1.2$), which restores transformation diversity without modifying the framework. 
StarCoder2-7B shows a more imbalanced distribution compared to other models, which is qualitatively consistent with the slower convergence reflected in the training dynamics~(\Cref{fig:training_llms}).

Model-level overhead remains consistent, exhibiting trends similar to those observed for the primary foundation model. This suggests that the computational characteristics of \theTool scale predictably across different LLM architectures.

\vspace{.4em}
\noindent
$\runtime{\text{{\bf \emph{Answer to RQ1. }}}}$ \theTool achieves strong performance on benchmark datasets and generalises effectively across different LLMs. Two-Level Monitoring significantly improves semantic validity while maintaining comparable generation behaviour and predictable computational overhead.

\begin{table}[t]
\centering
\caption{\small Complexity and computational cost across data scales}
\label{tab:scaling_cost_complexity}
\vspace{-1em}
\footnotesize
\setlength{\tabcolsep}{2pt} 
\begin{tabular}{lcccccccc}
\toprule
\multirow{2}{*}{\textbf{Pairs}}
& \multicolumn{2}{c}{\textbf{No Monitoring}}
& \multicolumn{2}{c}{\textbf{With Monitoring}} \\ 
\cmidrule(lr){2-3}\cmidrule(lr){4-5}
& {\bf Br.} / {\bf Msg.} & {\bf Time} (min) / {\bf Tok} (M)
& {\bf Br.} / {\bf Msg.} & {\bf Time} (min) / {\bf Tok} (M) \\
\midrule

0
 & 1.76 / 6.78 & 1578.16 / 8.5
 & 2.03 / 8.06 & 580.65 / 27.3 \\

602
 & 1.95 / 5.86 & 338.48 / 8.5
 & 2.35 / 6.73 & 1120.37 / 27.1 \\

9500
& 2.01 / 8.12 & 486.02 / 8.5
& 2.13 / 7.66 & 1575.50 / 27.3 \\

10800
& 1.83 / 8.06 & 335.69 / 8.6
& 1.93 / 8.54 & 1041.56 / 27.5 \\
\bottomrule
\end{tabular}
\vspace{-1.7em}
\end{table}

\subsection{RQ2. Impact of Training Data Scale on Performance}
\label{sec:rq_2}

To study how training data scale affects performance, we fine-tune models with LoRA on increasing numbers of 
$(\text{supertype}, \text{subtype})$ pairs. 
Due to variability in subtype generation, we approximate three regimes (small, large, and near-saturation) using $602$, 
$9,500$, and $10,800$ pairs, respectively, along with a no-fine-tuning baseline.

\Cref{fig:training_data_scaling} reveals that training data size significantly affects convergence behaviour, while transformation rule distributions and both syntactic and semantic validity vary accordingly, as shown in Fig.~\ref{fig:data_scaling_combined}\subref{fig:distribution_scaling} and~Fig.~\ref{fig:data_scaling_combined}\subref{fig:validity_scaling}.  Additional metrics and computational costs are reported in~\Cref{tab:scaling_cost_complexity}.

\vspace{.3em}
\noindent
{\bf No fine-tuning (0 pairs). }  Without fine-tuning, semantic validity is low (9.7\%), and generated subtypes are structurally simple, with limited transformation diversity and few reordering cases. Two-Level Monitoring improves validity substantially but does not address the lack of structural diversity, highlighting the need for fine-tuning to generate valid and diverse subtypes. 

\vspace{.3em}
\noindent
{\bf Small-scale (602 pairs). }  Fine-tuning yields significant performance gains. Pre-monitoring semantic validity increases from 9.7\% to 
31.2\%, while Two-Level Monitoring maintains validity at 85--90\%. Structural diversity also improves, with more frequent complex transformations. These results indicate that even limited training data provides meaningful improvements, although performance remains far from saturation.

\vspace{.3em}
\noindent
{\bf Large-scale (9,500 pairs). }  At this scale, performance reaches a near-optimal level. Pre-monitoring semantic validity approximately doubles relative to the 602-pair setting, reaching 98\% after Two-Level Monitoring. Structural diversity is well balanced, with broader coverage of complex transformations.  This is consistent with the model having largely converged, with limited scope for further improvement.

\vspace{.3em}
\noindent
{\bf Near-saturation (10,800 pairs). } Increasing the training size from 9,500 to 10,800 pairs yields only marginal changes (approximately ~1\%), indicating that performance has effectively saturated. This implies diminishing returns from further scaling and confirms that the model has reached a near-saturation regime.

\vspace{.4em}
\noindent
{\bf Computational Cost. } As shown in~\Cref{tab:scaling_cost_complexity}, computational cost exhibits minimal variation as the training dataset scales from 
602 to 10,800 pairs, incurring only minor differences in token consumption.
This shows that data scaling has limited impact on efficiency. The relative overhead introduced by Two-Level Monitoring remains consistent with the analysis in~\Cref{sec:result_Qwen2.5-Coder-7B}.  Overall, the framework demonstrates stable and predictable computational behaviour across different scales.

\vspace{.3em}
\noindent
$\runtime{\text{{\bf \emph{Answer to RQ2. }}}}$  Performance improves rapidly as training data increases, but saturates around 9,500 pairs, beyond which additional data yields only marginal gains. 

\vspace{-.2em}
\subsection{RQ3. Ablation Study of \theTool Components}
\label{sec:RQ_3}
\label{sec:rq_3}

\begin{figure*}[!t]
 \centering
\includegraphics[height=0.18\textheight,keepaspectratio]{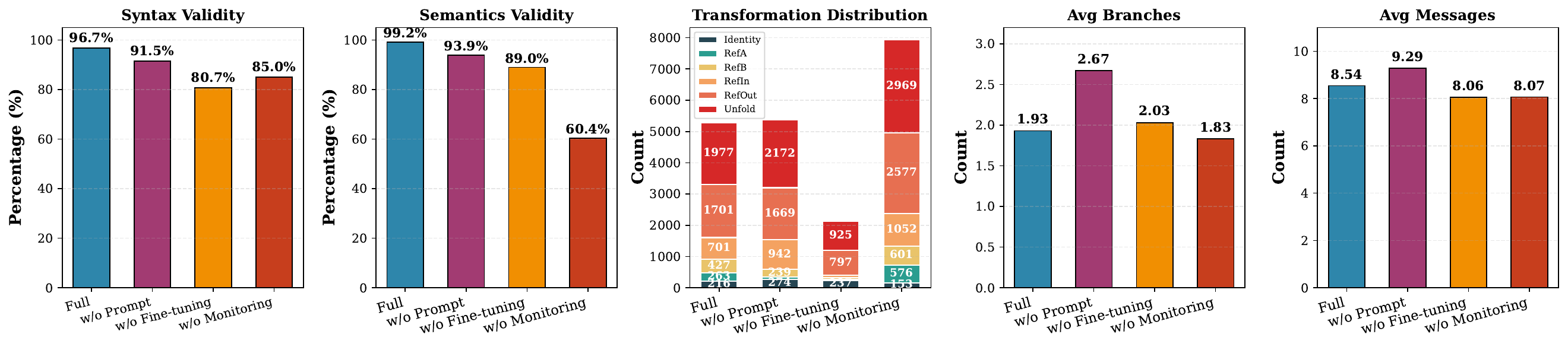}
\vspace{-2em}
\caption{\small Ablation results across metrics on Qwen2.5-Coder-7B}
\label{fig:ablation/ablation_visualization}
\end{figure*}

\begin{table*}[!t]
\centering
\caption{\small Semantic validity of diverse transformations across LLMs}
\label{tab:rule_scala}
\vspace{-1em}
\footnotesize
\setlength{\tabcolsep}{10pt}  
\begin{tabular}{lcccccc}
\toprule
& \multicolumn{3}{c}{{\bf Reordering}}
& \multicolumn{3}{c}{{\bf Variance}} \\
\cmidrule(lr){2-4} \cmidrule(lr){5-7}
{\bf Model} 
& {\bf \RefA} 
& {\bf \RefB}  
& {\bf Avg.} 
& {\bf \RefIn}  
& {\bf \RefOut}  
& {\bf Avg.}  \\
\midrule

\rowcolor{gray!15}
Qwen2.5-Coder-7B
& 261/263 (99.2\%) & 421/427 (98.6\%) & 673/681~(98.8\%) & 681/701~(97.1\%) & 1649/1701~(96.9\%) & 2167/2234~(97.0\%) \\

CodeLlama-7B 
& 230/231~(99.6\%) & 377/379~(99.5\%) & 606/609~(99.5\%) & 683/695~(98.3\%) & 554/568~(97.5\%) & 1100/1125~(97.8\%) \\

StarCoder2-7B 
& 210/211~(99.5\%) & 485/496~(97.8\%) & 682/694~(98.3\%) & 773/838~(92.2\%) & 1010/1159~(87.1\%) & 1695/1887~(89.8\%) \\

Qwen2.5-7B
& 213/213~(100\%) & 331/331~(100\%) & 544/544(~100\%) & 739/799~(92.5\%) & 1209/1301~(92.9\%) & 1789/1939~(92.3\%) \\

Qwen2.5-Coder-32B 
& 314/314~(100.0\%) & 433/439 (98.6\%) & 744/750 (99.2\%) & 973/1028 (94.6\%) & 1612/1731 (93.1\%) & 2378/2545 (93.4\%)  \\

\bottomrule
\end{tabular}
\vspace{-.7em}
\end{table*}

To understand the contribution of each component in \theTool, we conduct ablation studies on Qwen2.5-Coder-7B using three variants: \emph{w/o Prompt}, 
\emph{w/o Fine-tuning}, and \emph{w/o Two-Level Monitoring}. 
As the impact of fine-tuning has been analysed in RQ2~(\Cref{sec:rq_2}), we focus on prompting and monitoring. 
The results of the ablation study on validity, transformation rule distribution, and output complexity are illustrated in~\Cref{fig:ablation/ablation_visualization}.

\vspace{.3em}
\noindent
{\bf w/o Two-Level Monitoring. } This configuration exhibits the most severe degradation. Semantic validity drops to 60.4\%, considerably lower than all other configurations ($\geq 85\%$), highlighting that Two-Level Monitoring is essential for ensuring correctness.

\vspace{.3em}
\noindent
{\bf w/o Prompt. } 
Removing the BNF-style prompt does not substantially degrade validity, as both syntactic and semantic validity remain above 90\%. However, semantic validity decreases from 99.2\% to 93.9\%, indicating that prompting contributes to achieving near-perfect correctness. In contrast, structural diversity decreases more noticeably, with fewer reordering transformations, highlighting the importance of prompting for balanced transformation coverage.

\vspace{.3em}
\noindent
{\bf Full \theTool framework. } The complete system achieves the best overall performance, with 99.2\% semantic validity and balanced structural diversity.

\vspace{.4em}
\noindent
$\runtime{\text{{\bf \emph{Answer to RQ3. }}}}$  The components serve complementary roles: prompting enhances structural diversity, and Two-Level Monitoring ensures correctness. Combined with fine-tuning (RQ2), the full framework achieves the best overall performance. Removing any component degrades performance, with monitoring being most critical for validity and prompting for diversity. 

\subsection{RQ4. Robustness under Reordering and Variance Transformations}
\label{sec:rq_4}

\begin{table}[t]
\centering
\caption{\small Transformation preferences across two structural cases on Qwen2.5-Coder-7B}
\label{tab:case_rule_scala}
\vspace{-1em}
\footnotesize
\setlength{\tabcolsep}{1pt}
\begin{tabular}{@{}lcccccc@{}}
\toprule
& \multicolumn{3}{c}{\bf Reordering}
& \multicolumn{3}{c}{\bf Variance} \\
\cmidrule(lr){2-4} \cmidrule(lr){5-7}
{\bf Case}
& {\bf \RefA} 
& {\bf \RefB}  
& {\bf Avg.} 
& {\bf \RefIn} 
& {\bf \RefOut}  
& {\bf Avg.} \\
\midrule

\stT[1]
& 14/15 (93.3\%)  & 4/4 (100\%) & 18/19 (94.7\%) & 13/13 (100\%) & 32/33 (97.0\%) & 39/40 (97.5\%) \\

${\stT[1]}^{\!\looparrowright}$ 
& 19/19 (100\%)  & 8/8 (100\%) & 27/27 (100\%) & 16/16 (100\%) & 35/35 (100\%) & 46/46 (100\%) \\

\stT[2]
& 0 & 14/14 (100\%) & 14/14 (100\%) & 12/12 (100\%) & 38/38 (100\%) & 41/41(100\%) \\
\bottomrule
\end{tabular}
\vspace{-1.6em}
\end{table}

To evaluate the robustness of \theTool under different subtyping transformations, we analyse performance across two core categories: reordering (\RefA and \RefB) and variance (\RefIn and \RefOut). 
As shown in~\Cref{tab:rule_scala}, both achieve consistently high semantic validity across all evaluated models (89.8\%--100\%),  demonstrating strong robustness even for structurally complex reordering.

However, the generated subtype distribution is not balanced, with variance cases consistently outnumbering reordering ones.   
To investigate this, we introduce two representative structural cases: 

\vspace{.3em}
\centerline{\(
\small
\begin{array}{l}
\begin{array}{l}
\stT[1] = \stFmt{\mpFmt{\texttt{REC}\_\stFmt{\mathbf{X}}\_\texttt{OPEN}}  \,\roleFmt{p} ? {\stLabFmt{upd}}_{\roleP}; \roleFmt{q} ? {\stLabFmt{upd}}_{\roleQ};       
\mpFmt{\texttt{lbrace}}  \,\roleFmt{m} ! \stLabFmt{std};  \stFmt{\mpFmt{\texttt{REC}\_\stFmt{\mathbf{X}}\_\texttt{CLOSE}}},} 
\\
\kern17.15em \stFmt{\roleFmt{m} ! \stLabFmt{wtd};  \stFmt{\mpFmt{\texttt{REC}\_\stFmt{\mathbf{X}}\_\texttt{CLOSE}}}\,    \mpFmt{\texttt{rbrace}}}
\end{array}
\\[1em]
\begin{array}{l}
\stT[2] = \stFmt{\mpFmt{\texttt{REC}\_\stFmt{\mathbf{X}}\_\texttt{OPEN}}  \,\roleFmt{q} ? {\stLabFmt{upd}}_{\roleQ};       
\mpFmt{\texttt{lbrace}}  \,\roleFmt{m} ! \stLabFmt{std};  \stFmt{\mpFmt{\texttt{REC}\_\stFmt{\mathbf{X}}\_\texttt{CLOSE}}},}
\\
\kern13.8em \stFmt{\roleFmt{m} ! \stLabFmt{wtd};  \stFmt{\mpFmt{\texttt{REC}\_\stFmt{\mathbf{X}}\_\texttt{CLOSE}}}\,  \mpFmt{\texttt{rbrace}}}
\end{array}
\end{array}
\)}

\vspace{.3em}
\noindent
and generate their subtypes to analyse the resulting transformation distribution on Qwen2.5-Coder-7B. 
In addition, we consider a \emph{strengthening reordering} setting applied to \stT[1], where constraints are introduced to encourage reordering. The results are summarised in~\Cref{tab:case_rule_scala}, where ${\stT[1]}^{\!\looparrowright}$ denotes increased reordering.

For \stT[1], both \RefA and \RefB are allowed, yet the generated subtypes favour \RefA, indicating a bias toward transformations aligned with the prefix structure. For \stT[2], removing the initial receive action enables \RefB-style reordering; consequently,  \RefA disappears while \RefB  increases, confirming that both reordering types are supported when structurally enabled.

Across both cases, variance transformations exceed reordering by approximately two to three times, suggesting that they are easier to produce. While strengthening reordering constraints increases their proportion, variance remains dominant, indicating a persistent preference for simpler patterns.

\vspace{.4em}
\noindent
$\runtime{\text{{\bf \emph{Answer to RQ4. }}}}$  \theTool  maintains high semantic validity across both reordering and variance transformations, demonstrating strong robustness. However, the model exhibits a preference for simpler (variance) transformations, while reordering can be partially steered through constraints. 

\vspace{-0.1em}
\subsection{Error Analysis and Threats to Validity}
\label{sec:threat_error}

{\bf Syntax Validity. } 
LLMs occasionally produce syntactic errors, including mismatched parentheses, incomplete expressions (e.g., ending with $<$ or $;$), and unbound recursive variables. These errors reduce syntactic validity and are not fully eliminated by syntax normalisation during fine-tuning. This suggests that LLMs may be less reliable when handling mathematically structured formats compared to more common code-like syntax. Rather than introducing dedicated syntax correction techniques, we mitigate these errors indirectly: the two-level monitoring framework filters many invalid outputs, while multiple generations ensure a sufficient number of syntactically valid subtypes.

\vspace{.3em}
\noindent
{\bf Semantic Validity. } 
Ensuring semantic validity is challenging due to the undecidable nature of subtyping and the incompleteness of the subtyping checker used for validation~\cite{DBLP:conf/concur/BocchiK0025}. 
While accepted subtypes are guaranteed to be correct, rejected cases cannot be conclusively deemed incorrect, introducing uncertainty in evaluating borderline cases.

\vspace{.3em}
\noindent
{\bf Threats to Validity.  } 
The validity and diversity of generated subtypes depend on the characteristics of the underlying LLM. Although fine-tuning improves performance, generation quality remains influenced by the choice of model and may vary under different configurations.

In addition, our evaluation focuses exclusively on subtype generation. While consistent performance is observed across multiple LLMs, we do not assess generalisation to other generation tasks. Extending the approach beyond subtyping remains an important direction for future work.

\begin{table*}[t]
\caption{\small Performance and computational cost of prompting frontier models}
\label{tab:frontier_models}
\vspace{-1em}
\footnotesize
\begin{tabular}{lcccccccccccc}
\toprule
& \multicolumn{1}{c}{\bf Supertype-Level}
& \multicolumn{9}{c}{\bf Generated Subtypes}
& \multicolumn{2}{c}{\bf Cost} \\
\cmidrule(lr){2-2}
\cmidrule(lr){3-11}
\cmidrule(lr){12-13}
{\bf Model}
&  \cellcolor{gray!15}{{\bf Cov.} (\%)}
& {\bf Syn.} (\%)
& {\bf Sem.} (\%)
& {\bf Br.} / {\bf Msg.}
& {\bf \RefA}
& {\bf \RefB}
& {\bf \RefIn}
& {\bf \RefOut}
& {\bf \Unfold}
& {\bf \Identity}
& {\bf Time} (min)
& {\bf Tok} (M) \\
\midrule
GPT-5.5
&  \cellcolor{gray!15}{18}
& 94.59
& 99.29
& 1.78 / 7.30
& 18
& 2
& 49
& 13
& 50
& 52
& 59.47
& 0.5 \\

DeepSeek-V4-Pro
&  \cellcolor{gray!15}{4}
& 100.00
& 95.65
& 1.00 / 5.35
& 4
& 0
& 11
& 0
& 0
& 16
& 41.47
& 0.5 \\
\bottomrule
\end{tabular}
\label{tab:llm-comparison}
\end{table*}

\subsection{Comparison with Frontier Language Models}
\label{sec:comparison_gpt_deepseek}

To assess the necessity of a task-specific framework in the presence of modern frontier language models, 
we compare \theTool with GPT-5.5~\cite{openai2026gpt55} and DeepSeek-V4-Pro~\cite{deepseekai2026deepseekv4highlyefficientmilliontoken}, representative leading proprietary and 
open-weight models, respectively, in terms of both effectiveness and computational cost. 

Both models are evaluated on the benchmark using the \emph{same} prompts as the fine-tuned models under their  default inference configurations. \Cref{tab:frontier_models} summarises the results, including {\bf \emph{coverage}}. 

\vspace{.3em}
\noindent
{\bf Performance Analysis. }  
The key difference between frontier and fine-tuned models lies in their ability to generate subtypes across the benchmark. As shown in~\Cref{tab:frontier_models}, GPT-5.5 and DeepSeek-V4-Pro cover 18\% and 
4\% of benchmark supertypes, respectively. In contrast, all five fine-tuned models provide {\bf full} coverage.

Among the generated outputs, GPT-5.5 and DeepSeek-V4-Pro exhibit syntactic and semantic validity broadly consistent with those of the fine-tuned models (\Cref{tab:overall_results}). GPT-5.5 achieves 94.59\% syntactic validity and 99.29\% semantic validity, while DeepSeek-V4-Pro records 100.00\% and 95.65\%, respectively. GPT-5.5 additionally produces subtypes spanning all transformation categories and shows comparable branch and message counts.  However, these metrics need to be interpreted in the context of their substantially lower coverage.  
We further observe an overall tendency for covered supertypes to involve fewer message exchanges and simpler structures.

\vspace{.3em}
\noindent
{\bf Cost Analysis. }   
 Based on the results from~\Cref{tab:frontier_models,tab:training_llms_cost,tab:overall_results}, prompting frontier models requires fewer tokens and less time than the combined cost of fine-tuning and evaluating the models used in this work.  For the latter, the figures include both model training (\Cref{tab:training_llms_cost}) and benchmark generation (\Cref{tab:overall_results}), with training accounting for a modest fraction of the overall expenditure.  
Following~\Cref{sec:evaluation_rq_1}, token consumption remains the primary basis for comparison.
 The lower token usage of frontier models is partly attributable to their limited coverage and significantly fewer subtype candidates per covered supertype, averaging $8.22$ and $5.75$ for GPT-5.5 and DeepSeek-V4-Pro, respectively,  compared with $19.20$--$35.30$ for  the fine-tuned models. 

\vspace{.3em}
\noindent
{\bf Discussion. }  Taken together, these findings indicate that, although frontier models can generate valid subtype candidates at lower computational cost, they do not provide sufficient coverage for comprehensive subtype generation. The effectiveness gains achieved by the fine-tuned \theTool models, while maintaining high validity, therefore justify the need for this work.

\section{Related Work}
\label{sec:related}
{\bf Asynchronous Subtyping. }  
Asynchronous subtyping was initially shown to be sound and complete for binary session types (\ie involving two participants) in~\cite{ChenDSY17}. 
It was subsequently extended to multiparty session types~\cite{GPPSY2023}, with its formalisation and correctness proof mechanised in Rocq~\cite{tocl/10.1145/3815176}. For a comprehensive survey of asynchronous subtyping, see~\cite{DBLP:conf/ppdp/ChenDY24}.

Due to the undecidability of asynchronous subtyping~\cite{DBLP:journals/iandc/BravettiCZ17,DBLP:conf/fossacs/LangeY17}, even in the binary case, 
existing work focuses on developing sound, though necessarily incomplete, algorithms for subtyping verification. 
Approaches such as~\cite{DBLP:conf/concur/BravettiCLYZ19,DBLP:conf/tacas/BocchiKM24} provide practical checking procedures for binary session types, while~\cite{DBLP:conf/ppopp/CutnerYV22,DBLP:conf/concur/BocchiK0025} extend these ideas to multiparty settings based on a formal definition of asynchronous multiparty subtyping~\cite{GPPSY2023}, 
which precisely characterises safety-preserving type refinements, thereby guaranteeing correctness, including deadlock freedom. 
Benchmarks from~\cite{DBLP:conf/concur/BravettiCLYZ19,DBLP:conf/tacas/BocchiKM24,DBLP:conf/ppopp/CutnerYV22,DBLP:conf/concur/BocchiK0025} are adopted in the training and evaluation datasets of our framework, while the approach and checker in~\cite{DBLP:conf/concur/BocchiK0025} serve as the basis for our constrained generation and semantic validity checking. 

However, the automatic generation of valid asynchronous subtypes remains  unexplored in existing work.

\vspace{.3em}
\noindent
{\bf LLMs for Formal Specification.  } 
Transformer-based language models \cite{NIPS2017_3f5ee243}, such as Qwen-Coder~\cite{hui2024qwen25codertechnicalreport}, 
CodeLlama~\cite{DBLP:journals/corr/abs-2308-12950}, and StarCoder~\cite{DBLP:journals/tmlr/LiAZMKMMALCLZZW23}, 
have substantially advanced natural language tasks such as summarisation~\cite{pmlr-v149-chintagunta21a}, 
code generation~\cite{DBLP:journals/corr/abs-2107-03374}, and program synthesis~\cite{DBLP:journals/corr/abs-2108-07732}. 
Leveraging these strengths, a line of work~\cite{10.1007/978-981-96-0617-7_1,chen-etal-2023-nl2tl,DBLP:conf/cav/CoslerHMST23,DBLP:conf/aaai/FuggittiC23,sundarsingh2026conformalnl2ltltranslatingnaturallanguage,DBLP:conf/icfem/ZhaoTHSQY24,DBLP:conf/fmcad/MendozaHT24,DBLP:conf/kbse/MaWSLTQY25} investigates the translation of natural language into formal specifications, particularly temporal logic, to reduce the effort and error-proneness.  
Beyond specification generation, recent studies explore improving the reliability of LLM-based program synthesis by integrating formal reasoning techniques~\cite{DBLP:conf/nips/LinCHWLLSL24,wang2025pregussanalyzesspecifiesverifies}, including theorem proving, static analysis, and deductive verification.

These efforts demonstrate the complementary strengths of LLMs and formal methods in supporting the development of trustworthy software systems, motivating our exploration of integrating LLMs with the formal specification and verification of communication protocols. 

\vspace{.3em}
\noindent
{\bf Constrained Generation with LLMs.  }  Constrained generation aims to ensure that LLM outputs satisfy specified constraints and has mainly been studied in the context of constrained decoding. Syntactic constraints, particularly grammar-constrained decoding based on context-free grammars, have been extensively explored~\cite{DBLP:journals/pacmpl/Beurer-Kellner023,DBLP:conf/icml/Beurer-Kellner024,DBLP:conf/emnlp/GengJP023,park2025flexibleefficientgrammarconstraineddecoding,DBLP:journals/tmlr/UgareSKM025,DBLP:conf/nips/WangW0CSK23,willard2023efficientguidedgenerationlarge,DBLP:conf/iclr/PoesiaP00SMG22}. Simple context-sensitive features, such as indentation in \Python and scope markers in \Go, have also been incorporated~\cite{melcer2024constraineddecodingfillinthemiddlecode,DBLP:journals/tmlr/UgareSKM025}. 

In addition to syntactic constraints, recent work has investigated the enforcement of semantic constraints, such as type safety~\cite{DBLP:journals/pacmpl/MundlerHWSSV25,DBLP:journals/pacmpl/NagyZPD26}, as well as more general mechanisms, including monitors~\cite{DBLP:conf/nips/AgrawalKGLR23}, which enable user-defined constraint checking during decoding. These approaches inspire the design of our two-level monitoring strategy within constrained generation to enforce semantic correctness constraints of generated protocol refinements in \theTool. 

\vspace{.3em}
\noindent
{\bf LLM-based Communication Protocols. } The automatic generation of communication protocols among LLM agents has been explored, where emergent communication frameworks show that shared protocols can arise from decentralised interactions~\cite{DBLP:journals/corr/abs-2501-00226}, while alternative schemes, such as embedding-based protocols, enable richer information exchange~\cite{DBLP:conf/iclr/0001LY0LYP0Y24}. In addition, LLMs have been used to dynamically construct or adapt communication protocols at runtime~\cite{prakash2026ldpidentityawareprotocolmultiagent,DBLP:journals/corr/abs-2410-11905}. 
These approaches shift from static, human-designed protocols toward adaptive and learned communication paradigms; however, their underlying motivation differs from ours and is not grounded in formal specifications.

\section{Conclusion}
\label{sec:conclusion}
In this paper, we explored how large language models can be extended beyond syntactic code generation to provide  behavioural guarantees in communication protocols. By embedding formal constraints into the LLM-based generation process, the proposed framework \theTool enables the construction of protocol refinements that preserve properties such as deadlock freedom. 
The results demonstrate that integrating generative models with formal specification and verification techniques facilitates  correctness-preserving synthesis and highlights promising directions for applying LLMs to safety-critical software engineering tasks. 

In future work, we intend to extend the approach to more general formalisms, including choreographies, contract-based models, and temporal logics. We further plan to investigate verifier-guided iterative refinement techniques and the integration of \theTool into practical development workflows, including protocol evolution, interactive design assistance, and verification-aware code generation within existing toolchains.

\section*{Acknowledgments} 
We thank the reviewers for their detailed and constructive comments and suggestions. 
This work was partially supported by the EPSRC under grants EP/Z533749/1, EP/T006544/2, EP/V000462/1, EP/T014709/2, EP/X015955/1, EP/Y005244/1, and EP/N027833/2; 
Horizon EU TaRDIS 101093006 (UKRI No. 10066667); and the Advanced Research and Invention Agency (ARIA) Safeguarded AI.


\section*{Data Availability Statement}
The artifact supporting this work, including the implementation of \theTool, trained models, evaluation data, and evaluations of frontier language models, 
is publicly available via DOI: \url{https://doi.org/10.5281/zenodo.21737877}. The repository contains all source code, datasets, and instructions required to reproduce the experiments and results reported in this paper.







\bibliography{references,popl19,llm}

\iftoggle{full}
{
  \newpage
  \appendix
  \label{appendix}
}{
}

\end{document}
\endinput